\documentclass{aa}
\usepackage{graphicx}
\usepackage{xspace} 
\usepackage{multirow}
\usepackage{dcolumn}
\newcolumntype{d}[1]{D{.}{.}{#1}}
\usepackage{hyperref}
\usepackage{xcolor}
\usepackage{soul}
\hypersetup{
    colorlinks,
    linkcolor={blue!},
    citecolor={blue!},
    urlcolor={blue!},
}
\let\oldpageref\pageref
\renewcommand{\pageref}{\oldpageref*}
\usepackage{txfonts}
\usepackage{amsmath}
\usepackage{natbib}
\usepackage{lscape}
\usepackage{multicol}
\usepackage[version=4]{mhchem}
\usepackage{booktabs}
\usepackage{xcolor}
\usepackage[normalem]{ulem}
\newcommand{\meng}[1]{#1}

\newcommand{\modb}[1]{\iffalse #1 \fi}
\newcommand{\moda}[1]{{#1}}

\newcommand{\LEt}[1]{}
\usepackage[detect-all,load=abbr)]{siunitx}
\newcommand{\hii}{H{\sc ii }}
\newcommand{\kms}{ ${\rm km\ s^{-1}}$ }

\begin{document} 

\title{The physical and chemical structure of Sagittarius\,B2}
\titlerunning{UCH{\sc ii} regions in Sgr\,B2}
\subtitle{VI. UCH{\sc ii} regions in Sgr\,B2}
\author{F.~Meng\inst{1,2},
 {\'A}.~S{\'a}nchez-Monge\inst{2},
 P.~Schilke\inst{2},
 A.~Ginsburg\inst{3},
 C.~DePree\inst{4,5},
 N.~Budaiev\inst{3},
 D.~Jeff\inst{3},
 A.~Schmiedeke\inst{6},
 A.~Schw{\"o}rer\inst{2},
 V.~S.~Veena\inst{7},
\and
Th.~M{\"o}ller\inst{2}
          }
\authorrunning{F. Meng et al.}
\institute{University of Chinese Academy of Sciences, Beijing 100049, People's Republic of China\\
           \email{mengfanyi@ucas.ac.cn, meng@ph1.uni-koeln.de}
           \and
           I.\ Physikalisches Institut, Universit\"at zu K\"oln, Z\"ulpicher Str.\ 77, D-50937 K\"oln, Germany
           \and
           Department of Astronomy, University of Florida, PO Box 112055, USA
           \and 
           NRAO, 520 Edgemont Rd, Charlottesville, VA, USA
           \and
           Agnes Scott College, 141 E. College Ave., Decatur, GA 30030, USA
           \and
           Max Planck Institute for Extraterrestrial Physics, Giessenbachstrasse 1, D-85748 Garching, Germany
           \and 
           Max Planck Institute for Radio Astronomy, Auf dem H\"{u}gel 69, D-53121 Bonn, Germany
           }

\date{Received ; accepted }
 
\abstract{
\LEt{ General notes. A) You show a preference for US English language conventions, and I have edited accordingly throughout. B1) A\&A uses the past tense to describe the specific methods used in a paper and the present tense to describe general methods and recent findings (within the past ten or so years). Please make sure my edits are accurate in this respect throughout the paper. See Sect. 6 of the Language Guide https://www.aanda.org/for-authors/language-editing/6-verb-tenses B2) This means that the specific steps you took for this particular research should be in the past simple: We used, We extrapolated, The observation was performed, After this change we found,  etc. Descriptions of methodology, universal truths, constants, or conclusions should be in the present simple: The first step in the process is, Water boils at 100^{\circ}$C, We find, etc. }\\

The giant molecular cloud Sagittarius B2 (hereafter Sgr\,B2) is the most massive region with ongoing high-mass star formation in the Galaxy. Two
ultra-compact \hii (UC\hii) regions were identified in Sgr\,B2's central hot cores, Sgr\,B2(M) and Sgr\,B2(N).
} 
{
Our aim is  to characterize the properties of the \hii regions in the entire Sgr\,B2 cloud. 
Comparing the \hii regions and the dust cores, we aim to depict the evolutionary stages of different parts of Sgr\,B2.
} 
{
We use the Very Large Array in its A, CnB, and D configurations, and in the frequency band C ($\sim$6\,GHz) to observe the whole Sgr\,B2 complex. 
Using ancillary VLA data at 22.4\,GHz and  ALMA data at 96\,GHz, we calculated the physical parameters of the UC\hii regions and their dense gas environment.
} 
{
We identify 54 UC\hii regions in the 6\,GHz image, 39 of which are also detected at 22.4\,GHz. Eight of the 54 UC\hii regions are newly discovered.
The UC\hii regions have radii between $0.006\,{\rm pc}$ and $0.04\,{\rm pc}$, and have emission measure between $10^{6}\,{\rm pc\,cm^{-6}}$ and $10^{9}\,{\rm pc\,cm^{-6}}$. 
The UC\hii regions are ionized by \meng{stars of  types} from B0.5 to O6. 
We found a typical   gas density of $\sim10^6-10^9\,{\rm cm^{-3}}$ around the UC\hii regions. The pressure of the UC\hii regions and the dense gas surrounding them are comparable.
The expansion  timescale of these UC\hii regions is determined to be $\sim10^4-10^5\,{\rm yr}$.  
The percentage of the dust cores that are associated with \hii regions are 33\%, 73\%, 4\%, and 1\% for Sgr\,B2(N), Sgr\,B2(M), Sgr\,B2(S), and Sgr\,B2(DS), respectively. 
Two-thirds of the dust cores in Sgr\,B2(DS) are associated with outflows.
} 
{
The electron densities of the UC\hii regions we identified are in agreement with that of typical UC\hii regions, while the radii are smaller than those of the typical UC\hii regions. 
The dust cores in Sgr\,B2(M) are more evolved than  in Sgr\,B2(N). 
The dust cores in Sgr\,B2(DS) are younger than   in Sgr\,B2(M) or Sgr\,B2(N).
}


\maketitle

\section{Introduction} 
\label{sec:introduction}

  The giant molecular cloud Sagittarius B2 (Sgr\,B2) is the most massive ($\sim 10^7\,M_{\odot}$) region with ongoing high-mass star formation in the Galaxy \citep[see e.g.,][]{Goldsmith:1990aa}. 
  Sgr\,B2 has a higher density ($>10^5\rm\,cm^{-3}$) and dust temperature ($\gtrsim$50--70\,K) compared to other star forming regions in the Galactic plane \citep[see e.g.,][]{Ginsburg:2016aa,Schmiedeke:2016uc,Sanchez-Monge:2017vt}. 
  Sgr\,B2 is located at a distance of $8.34\pm0.16$\,pc, and only $\sim$100\,pc in projection from the Galactic center \citep{Reid:2014aa}\footnote{
    A new distance to the Galactic center of $8.127 \pm 0.031$~kpc has been measured \citep{Gravity-Collaboration:2018aa}. For consistency with the papers published within the same series of studies of Sgr\,B2, we use the distance reported by \citet{Reid:2014aa}. 
    }. 
  These features make Sgr\,B2 an excellent case to study high-mass star formation in an extreme  high-pressure environment. Such an environment resembles nearby starburst galaxies \citep{Leroy:2018to}. 
  Understanding the structure of the Sgr\,B2 molecular cloud complex is necessary to comprehend the most massive star forming region in our Galaxy, which at the same time provides a unique opportunity to study in detail the nearest counterpart of the extreme environments that dominate star formation in the Universe  \citep[see, e.g.,][]{Kruijssen:2013vk,Henshaw:2022vl}.
  This paper continues our series of studies on Sgr\,B2 \citep{Schmiedeke:2016uc,Sanchez-Monge:2017vt,Pols:2018aa,Schworer:2019aa,Meng:2019aa}.
  In \citet{Meng:2019aa} we presented the observations of Sgr\,B2(DS), which is a part of Sgr\,B2 giant cloud, and analyzed the physical properties of the \meng{non-thermal} emission within it.
  In this work we study the whole Sgr\,B2 region.

  In the central $2\,{\rm pc}$ of Sgr\,B2 there are the two well-known and well-studied hot cores Sgr\,B2(N) and Sgr\,B2(M) \citep[see, e.g.,][]{Schmiedeke:2016uc,Sanchez-Monge:2017vt}, which contain at least 70 high-mass stars with spectral types from O5 to B0 \citep[see, e.g.,][]{Gaume:1995aa,De-Pree:1998aa,De-Pree:2014aa}.  
  Surrounding the two hot cores, there is a larger envelope (hereafter \emph{the envelope}) with a radius of 20~pc that contains more than 99\% of the total mass of Sgr\,B2 \citep{Schmiedeke:2016uc}.
  Along with the active high-mass star forming activity discovered in Sgr\,B2(N) and Sgr\,B2(M), hints of star formation happening in the envelope are also revealed. 
  \citet{Ginsburg:2018wo}, with ALMA at 3~mm, revealed 271 high-mass protostellar cores distributed throughout the entire Sgr\,B2 region, including the envelope. 
  The luminosities of these dust cores suggest that they must contain objects with stellar masses higher than 8\,$M_\odot$. 

  Due to the high extinction in the infrared bands toward Sgr\,B2 \citep[see][]{Meng:2019aa}, there is no direct evidence of the existence of high-mass stars embedded in the dust cores detected by \citet{Ginsburg:2018wo}.\LEt{ ok like so? if you say "missing" it means that you had the information in the past, but now you cannot find it. } 
  However, since high-mass stars ionize the neutral material surrounding them, the presence and properties of the associated \hii regions reflect the evolutionary stages of these dust cores \citep[see, e.g.,][]{Gonzalez-Aviles:2005aa,Breen:2010aa}. 
  Additionally, since the free-free emission from \hii regions may extend from centimeter to millimeter wavelengths in the spectral domain \citep[see, e.g.,][]{Sanchez-Monge:2013ab}, measuring the luminosities of the associated \hii regions can help us better constrain the luminosities of the dust cores. 
  Therefore, to further characterize the evolutionary stages and physical properties of these dust cores, we     investigate the possible \hii regions associated with them. 

  The \hii regions in Sgr\,B2 were targeted by several previous studies. 
  \citet{Mehringer:1993aa} observed the entire Sgr\,B2 with VLA in the  20, 6, and 3.6\,cm bands and identified 15 \hii regions. 
  The resolutions range from  $\sim$20\arcsec\ to  $\sim$3\arcsec\ when the wavelength changes from 20 to 3.6\,cm, which correspond to the range 0.8--0.12\,pc. 
  The 15 \hii regions, except two unresolved cases, all have sizes $>2$\arcsec. 
  Since the 271 dust cores may contain newly formed high-mass stars \citep{Ginsburg:2018wo}, the associated \hii regions may be ultra-compact \hii (UCH{\sc ii}) and hyper-compact \hii (HCH{\sc ii}) regions,  which typically have sizes from $\sim 0.03$\,pc to $\sim 0.1$\, pc \citep[see, e.g.,][]{Kurtz:2002aa,Gonzalez-Aviles:2005aa,Kurtz:2005aa,Breen:2010aa}. 
  Thus, the resolution of \citet{Mehringer:1993aa} is not sufficient to resolve the UCH{\sc ii} and HCH{\sc ii} regions. 
  \citet{Gaume:1990aa,Gaume:1995aa} \LEt{ In the main text (i.e., not  in parentheses) use "and" between two references, not ";". Three or more references must be separated by commas and the last reference set off with "and": ref1, ref2, and ref3. Please check for this throughout.
 }observed Sgr\,B2(N) and Sgr\,B2(M) at 7\,mm and 1.3\,cm and achieved resolutions of 0.065\arcsec\ and 0.25\arcsec, respectively. 
  \citet{Rolffs:2011ty} observed Sgr\,B2(N) and Sgr\,B2(M) in 40\,GHz with a resolution of 0.1\arcsec. 
  Unfortunately, these high-resolution observations do not cover the entire envelope.
  For example, Sgr\,B2(DS), where  $\sim$80 of dust cores reside \citep[see, e.g.,][]{Ginsburg:2018wo,Meng:2019aa}, is left out of these observations. 
  \citet{LaRosa:2000wr,Law:2008tl,Law:2008vw}  also observed the entire Sgr\,B2 at  centimeter wavelengths, but the  resolutions were not high enough to study the UCH{\sc ii} and HCH{\sc ii} regions. 

  In this paper we present Very Large Array (VLA) observations of the entire Sgr\,B2 cloud in the frequency regime 4--8~GHz, with configurations A, BnC, and D. 
  The high resolution ($\lesssim 0.01$\, pc) and large spatial coverage ($\sim 20$\,pc) of our data sets make a systemic and complete study of the UC\hii and HC\hii regions in Sgr\,B2 possible.
  We also include analysis of the 3\,mm image \citep{Ginsburg:2018wo} as well as the newly acquired \ce{SiO}\,(5--4) data, both of which were observed with  the Atacama Large Millimeter/submillimeter Array (ALMA).
  Thus, we can disentangle the contributions of ionized gas and dust at millimeter wavelengths and better constrain the evolutionary stages of the dust cores.

  This paper is organized as follows.
  In Sect.~2 we describe the observations and the data reduction process. 
  In Sect.~3 we present the results.
  In Sect.~4 we discuss the results. 
  Finally, we summarize this paper in Sect.~6.

\section{Observations and data reduction} 
\label{sec:observations_and_data_reduction}

  We \moda{used} the VLA in \moda{its} A, CnB, and D configurations to observe the entire Sgr\,B2 complex in frequency band C (4--8~GHz). In the following we call this band  6\,GHz.
  The observations with the CnB and D configurations were described in \citet{Meng:2019aa}. 
  The observations with the A configuration were conducted from October 1 to 12, 2016 (project 16B-031, PI: F.\  Meng).
  We used 64 spectral windows with a bandwidth of 128~MHz each. 
  Mosaic mode was used, with ten pointings for C band. 
  The primary beam of each pointing is 7.5\arcmin. 
  Quasar 3C286 was used as the flux and bandpass calibrator, the SED of which is $S_\nu = 5.059\pm0.021\ {\rm Jy} \times (S/8.435\ {\rm GHz})^{-0.46}$ from 0.5 to 50~GHz \citep{Perley:2013aa}.
  Quasar J1820-2528, whose flux  is 1.3~Jy in the C Band, was used as phase calibrator.
  The calibration was done using the standard VLA pipelines provided by the NRAO\footnote{The National Radio Astronomy Observatory is a facility of the National Science Foundation operated under cooperative agreement by Associated Universities, Inc.}. 

\begin{figure*}[ht]
          \begin{center}
          \includegraphics[width=0.85\textwidth]{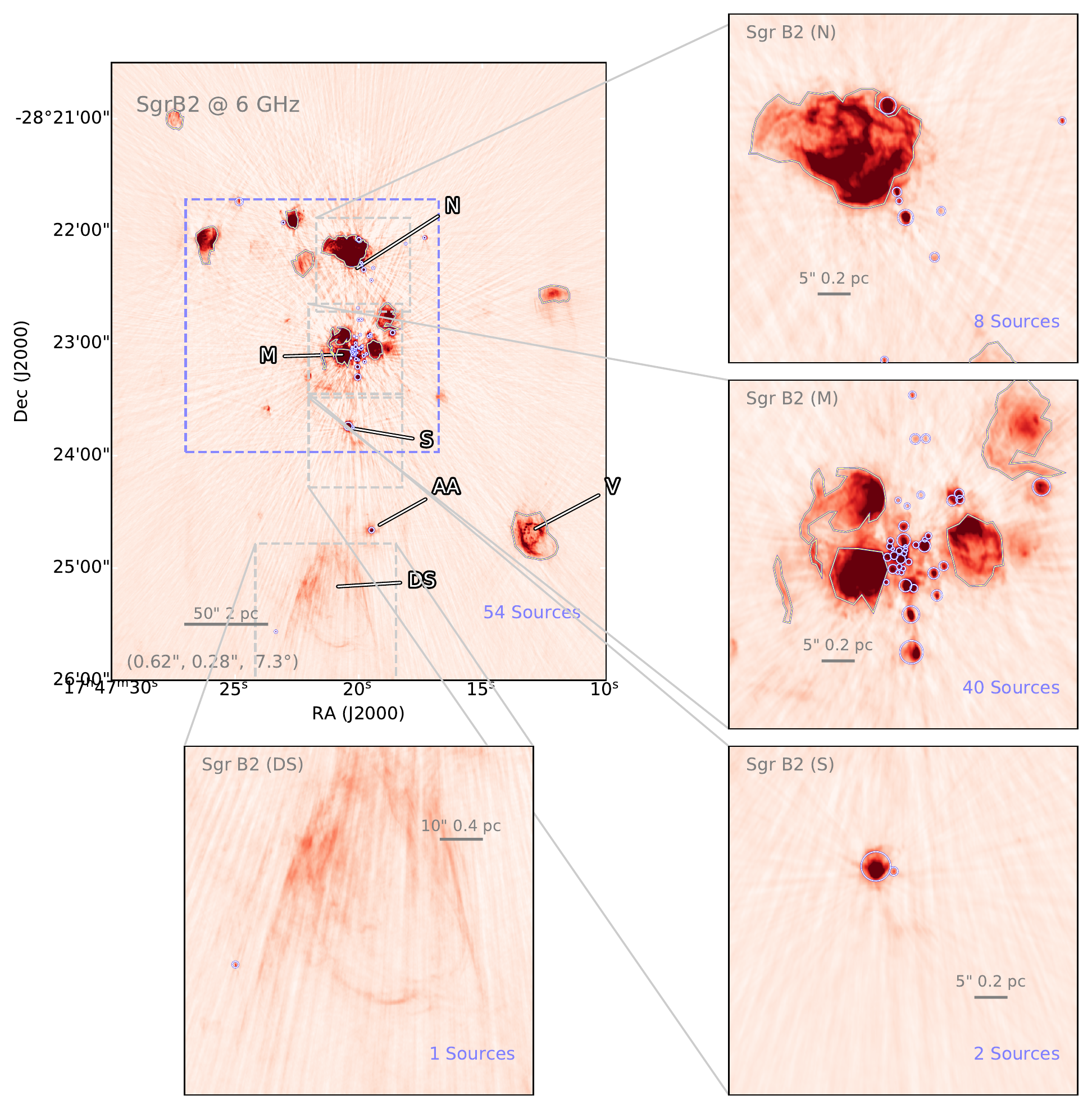}
          \caption[Cores in 6\,cm image]{Sources  identified in the  6\,cm image. Those objects shown as  white and blue circles have  radii of $r_{\rm obs6}$ (see Table\,\ref{t:coreparams}).\LEt{ have I interpreted correctly? }  Notable \hii regions are indicated. The \hii regions that are masked out for core identification are shown as white contours. 
           The beam size is given in  the lower left corner of the main panel $(\theta_{\rm maj}, \theta_{\rm min}, {\rm PA})$. The spatial coverage of the 22.4\,GHz image is shown  as a blue dashed box.
           }
          \label{f:coresoverallvla}
          \end{center}
  \end{figure*}

  Calibration and imaging were done in Common Astronomy Software Applications \citep[CASA 4.7.2][]{McMullin:2007aa}.
  The details of the  data processing  from the  CnB and D configurations are described in \citet{Meng:2019aa}.
  The A configuration data were originally taken every 2\,s. To shorten the time of processing, we applied \texttt{timebin} in CASA to the measurement sets, which averaged the data taken within 10\,s into one data point.
  All the pointings of the mosaic in each band were primary beam corrected and the mosaic was imaged using the CASA task \texttt{tclean}.
  With a robust factor of 0, the image of C band has a synthesized beam of 0.62\arcsec$\times$0.28\arcsec, with a position angle ${\rm (PA)}$ of $7.27^\circ$.
  The PA is defined  positive north to east.\LEt{ The PA is defined as positive from north to east. ? }
  To mitigate the spatial filtering effect of the A configuration image, we applied the \texttt{feather} algorithm to combine the images from A configuration and     from CnB and D configurations. The combined image of the three configurations is shown in Fig.\,\ref{f:coresoverallvla}.
  The combined image has a resolution identical to that of the A configuration images, while being sensitive to spatial scales up to $\sim240$\arcsec.
  The sensitivities of the observations are described in Sect.~\ref{sec:results}. 

  Ancillary data include the 22.4 GHz data, the 3\,mm continuum, and \ce{SiO}\,(5--4) data. 
  The observations of the 22.4 GHz data are described in \citet{Gaume:1995aa}, with a resolution of 0.27\arcsec$\times$0.23\arcsec\ (${\rm PA} = 70^\circ$). The root mean square (RMS) noise is 0.38 mJy/beam. The spatial coverage of the 22.4\,GHz data is shown as the blue dashed box in Fig.\,\ref{f:coresoverallvla}.
  The 96\,GHz continuum data covers the  frequency range from 89.5 to 103.3\,GHz. 
  The image in 96\,GHz has a resolution of 0.54\arcsec$\times$0.46\arcsec\ (${\rm PA}=68.31^\circ$), the observational details of which are described in \citet{Ginsburg:2018wo}. Unlike the 22.4\,GHz image, the 96\,GHz image covers the entire area shown in Fig.\,\ref{f:coresoverallvla}.
  The 22.4\,GHz observations were performed 27 years prior to the 6\,GHz and 96\,GHz observations. If we assume that Sgr\,B2 has a proper motion of $\sim 50$\kms \citep{Henshaw:2016ug}, the corresponding position shift in 27 years is 0.04\arcsec, which is only $\sim$10\% of the beam sizes of our images. 
  In addition, we did not find any visible shift in the  positions of the compact sources positions between the 6 and 22.4\,GHz images. Thus, we do not take the astrometry difference into account when analyzing the 6 and 22.4\,GHz data in the following sections.
  The SiO~(5--4) emission was observed with ALMA (Project 2017.1.00114.S, P.I. A. Ginsburg) and has a resolution of 0.35\arcsec $\times$ 0.24\arcsec, with \meng{PA}\LEt{ is this the same as PA, which you introduced above? Please make sure the abbreviations are consistent throughout. } of $-80^{\circ}$, and spectral resolution of 1.35~km~s$^{-1}$. 
  For the details of the observation and data reduction, see Jeff et al. (in prep.). 
  The typical RMS of the SiO image is 0.9~mJy/beam (0.3\,K). The observation covers Sgr\,B2(S) and the eastern part of Sgr\,B2(DS). 

 \section{Results} 
 \label{sec:results}

  In this section we present the image of Sgr\,B2 at 6\,GHz and the UCH{\sc ii} regions identified in it. 
  For the \hii regions, we calculated their actual sizes and physical properties using the observations at   6\,GHz and at 22.4\,GHz. With the 96\,GHz data we characterized the properties of the dense gas that the UC\hii regions reside in.

  \subsection{Observed parameters of the 6 GHz sources} 
  \label{sub:observed_parameters_of_the_6_ghz_cores}
  

    Figure\,\ref{f:coresoverallvla} displays the image of Sgr\,B2 at 6\,GHz, where the known  large-scale \hii regions are denoted  N, M, S, AA, DS, and V, following the nomenclature of \citet{Mehringer:1992aa,Mehringer:1993aa,Ginsburg:2018wo,Meng:2019aa}. 
    The maps at 22.4\,GHz and 96\,GHz are presented by \cite{Gaume:1995aa} and \cite{Ginsburg:2018wo}, respectively. 
    Our  aim is to study UC\hii regions and since the 22.4\,GHz image has non-complete spatial coverage, we performed compact source identification at the 6\,GHz image only. 
    We used the automatic source extracting algorithm SExtractor \citep{Bertin:1996aa}, which allows us to identify the location of bright compact sources throughout the map.
    We note that this algorithm also includes a large fraction of artificial sources due to some large-scale artifacts visible in the final image (see the  zoomed-in images in Fig.~\ref{f:coresoverallvla}).
    The variable noise (see Fig.\,\ref{f:rmsmaps}) also has an effect on the exclusion of certain sources with the automatic algorithms.
    After cross-checking the automatically produced lists of sources at 6\,GHz and 22.4\,GHz, we made the final catalog of continuum sources at 6\,GHz by excluding or adding sources by visual inspection (in a similar way to the approach followed in \citealt{Ginsburg:2018wo} for the ALMA 96\,GHz data).\LEt{ The slash "/" is used for ratios and some instrument pairings. Please check my changes throughout: you can substitute "and", "or", "and/or", or a double hyphen (which can be used to indicate dual nature: Hertzprung--Russell diagram).}
    Since we focus on compact sources, extended emission larger than 5\arcsec\ or 0.2\,pc (see  contours in Fig.\,\ref{f:coresoverallvla}) are excluded in the catalog  and the following analysis.
    In total, 54 compact sources are identified throughout the entire Sgr\,B2 cloud (see Table\,\ref{t:coreparams}), 8 of which are newly identified. Of these 54 cores, 8 are identified in Sgr\,B2(N), 40 in Sgr\,B2(M), 2 in Sgr\,B2(S), and 1 in Sgr\,B2(DS).  
    All   54 compact sources are covered by the 22.4\,GHz image, except core 1.

    For each of the 54 compact sources we define a minimal circle that can include as much as the total flux density of it at 6\,GHz.\footnote{
    \LEt{ Numerous  discursive footnotes, especially those that have a direct bearing on the information in the paper, should be avoided. Please move the footnotes to the main text whenever possible.
I think this  should be in the main text; I have made the necessary changes.   }This is done by expanding the circle as long as the flux $S_{\rm total}$ within the circle increases monotonically. When $S_{\rm total}$ remains constant or (due to the ``negative bowl'' in the artifacts) decreases, the corresponding circle is called the  minimum circle.
    }
    \LEt{ perhaps like so: ... we find the    smallest circle   that includes the greatest amount of  total flux density at 6 GHz   }The observed radius of the core and the flux density within the core are denoted   $r_{\rm obs6}$ and $S_{\rm 6}$, respectively. 
    For these 54 sources we followed the same photometry procedure at the  22.4\,GHz and 96\,GHz bands, and obtained $r_{\rm obs22}$, $S_{\rm 22}$, $r_{\rm obs96}$, and $S_{\rm 96}$ that correspond to the observed radii and flux densities. 
    The observed radius and flux density in three bands of all the 54 compact sources are listed in Table\,\ref{t:coreparams}.

    To match the sources across the three images (6, 22.4, and 96\,GHz), we defined that if two sources in two images have a distance (distance between their centers) shorter than either of their radii, they are associated with each other.\LEt{ ... we define two sources in two images as associated with each other if the distance between their centers is smaller than either of their radii.    } Afterward, we manually adjust few of the matched sources in the crowed region in Sgr\,B2(M).
    For all the compact sources at 6\,GHz, emission in the 96\,GHz band is also detected.
    Among the 15 sources that have no $S_{\rm 22}$, 14 are without reliable 22.4\,GHz detection ($S_{22}<3{\rm RMS}$), and one (\#1 in Tab.\,\ref{t:coreparams}) is not spatially covered by the 22.4\,GHz image.
    All the 22.4\,GHz sources\footnote{
    In \citet{Gaume:1995aa}, extended \hii regions are also included in the catalog. We only compare the compact sources ($r<5$\arcsec)   in their catalog with ours.}
    reported by \citet{Gaume:1995aa} are associated with the 6\,GHz compact sources.

    We compared the spatial association of the 54 sources with the 271 dust cores identified at 96\,GHz by \citet{Ginsburg:2018wo}.
    Although there is no one-to-one correspondence between the 271 dust cores and the associated dust emission of the 54 compact sources, we can still conclude that    at least 217 (80\%) dust cores are not associated with compact radio sources that are detectable with our sensitivity. 

    \begin{figure}[ht]
            \begin{center}
            \includegraphics[width=0.4\textwidth]{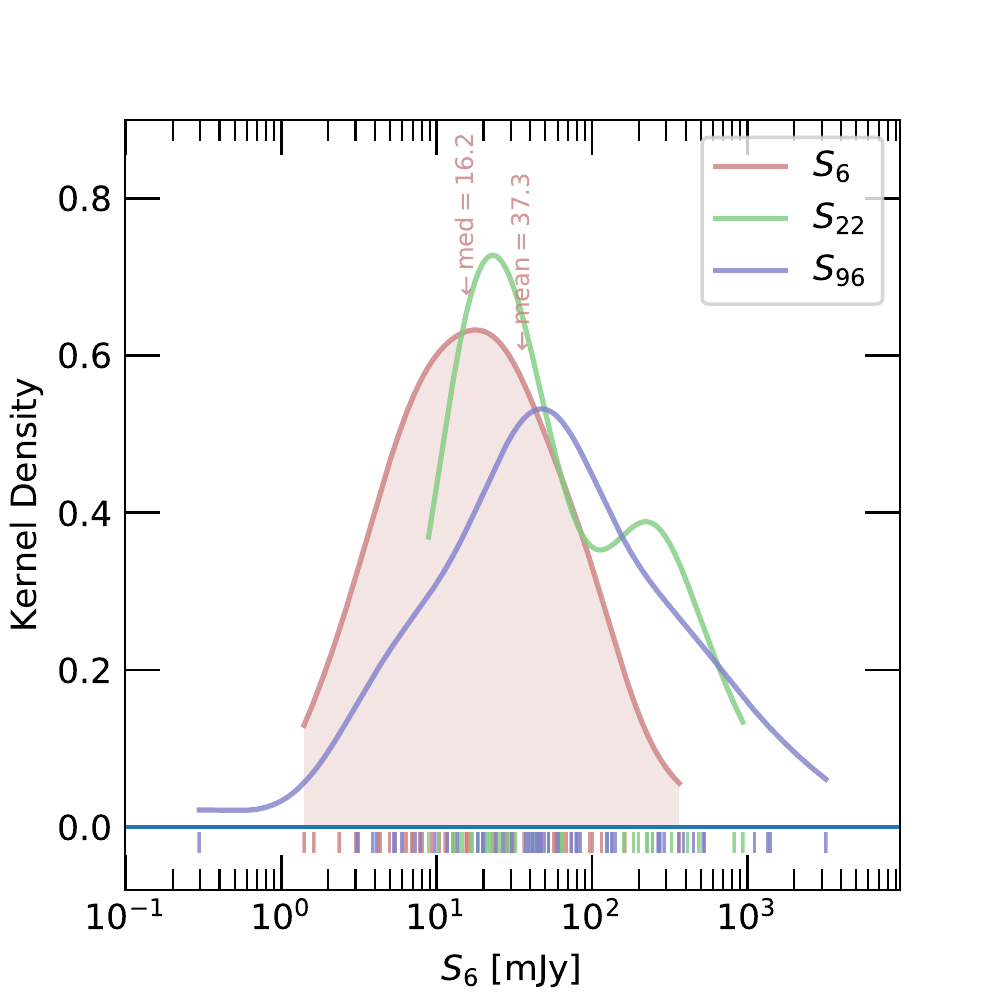}
            \caption[Distribution of $S_{\rm 6}$]{Probability distribution of $S_{\rm 6}$ (pink), plotted as a kernel density estimation. Mean and median values are indicated. The distributions of $S_{\rm 22}$ (green) and $S_{\rm 96}$ (blue) are also shown.
             }
            \label{f:s_cband_distribution}
            \end{center}
    \end{figure}

    Most of the compact sources have $r_{\rm obs6}$ from 0.5\arcsec\ to 1\arcsec, which is comparable to the beam size at 6\,GHz.  
    Even though  the 22.4\,GHz image has higher resolution ($\sim 0.25^{\prime \prime}$), the measured $r_{\rm obs22}$ are still not significantly larger than the beam size at 22.4\,GHz. 
    Therefore, most of the compact sources are not well resolved; in other words,  $r_{\rm obs6}$ or $r_{\rm obs22}$ cannot accurately represent the actual radii of most of the compact sources.
    The 96\,GHz compact sources, as indicated by \citet{Ginsburg:2018wo}, are also not resolved.
    The flux densities of the compact sources are distributed in a wide range; 
    $S_{\rm 6}$ ranges from $\sim 3$\,mJy to $\sim 300$\,mJy at 6\,GHz, as shown in Fig.\,\ref{f:s_cband_distribution}. The mean and median of $S_{\rm 6}$ are 37.3\,mJy and 16.2\,mJy, respectively. 

    To test whether the nature of the emission at 6 and 22.4\,GHz is free-free emission, we interpolated $S_{\rm 6}$ and $S_{\rm 22}$ to derive a spectral index $\alpha_{\rm 6-22}$ for each of the 39 sources that have 22.4\,GHz detection, see the last column of Table\,\ref{t:coreparams}.
    All   39 sources have $\alpha_{\rm 6-22}>-0.1$, which suggest that their emission at centimeter wavelengths is very probably dominated by thermal free-free emission from ionized \meng{gas} \LEt{ gas? } \citep[see, e.g.,][]{Sanchez-Monge:2013ab}. For the  15 sources without 22.4\,GHz detections, we find that their $S_{\rm 6}$ are not higher than the values of the other sources. If we treat their emission at 6\,GHz as free-free as well
    \footnote{For single-dish images, the RMS around a certain null detection spot could be used to derive an upper limit of the possible signal, which in this paper can be translated into an upper limit of $\alpha_{\rm 6-22}$; however, because the artifacts in the 22.4\,GHz image contain negative bowls of interferometric images, we do not trust the upper limits derived from the 22.4\,GHz RMS map. }, the derived physical parameters are also within the same ranges as the other sources (see Section\,\ref{sub:cores_in_vla}). Thus, we treat all   54 sources as thermal free-free sources in the following analysis.

  \subsection{Physical parameters of the H{\sc ii} regions} 
  \label{sub:cores_in_vla}

    As we show in Sect.\,\ref{sub:observed_parameters_of_the_6_ghz_cores}, most of the compact sources at 6 and 22.4\,GHz are not well resolved.
    Therefore, we first need to   determine their actual sizes.
    The observed flux density ($S_{\rm 6}$ and $S_{\rm 22}$) of an \hii region is related to its actual size ($r_{\rm calc}$,  which is called calculated radius in this paper), its electron temperature ($T_{\rm e}$), and its emission measure (EM) as \citep[see, e.g.,][]{wilson2013tools}
    \begin{equation}
    \label{eq:s-tau}
      \frac{S_\nu}{\rm mJy}= 8.183\times10^{-4} 
      \left(\frac{r_{\rm calc}}{\rm arcsec}\right)^{2}
      \left(\frac{\nu}{\rm GHz}\right)^{2}
      \left(\frac{T_{\rm e}}{\rm K}\right) (1-e^{-\tau_{\nu}}),
    \end{equation}
    where the optical depth $\tau_\nu$ is
    \begin{equation}
    \label{eq:tau-em}
      \tau_\nu = 8.235\times10^{-2}
      \left(\frac{\nu}{\rm GHz}\right)^{-2.1}
      \left(\frac{T_{\rm e}}{\rm K}\right)^{-1.35} 
      \left(\frac{\rm EM}{\rm pc\,cm^{-6}}\right).
    \end{equation}

    In our case the frequency $\nu$ is 6\,GHz or 22.4\,GHz.
    We assume that   $T_{\rm e}$ is $10^4$\,K, following the values given by \cite{Mehringer:1993aa} for Sgr\,B2(N) and Sgr\,B2(M), considering that most of the compact sources are in these two regions.
    Thus, with the two independent flux density measurements, $S_{6}$ and $S_{22}$, we solve Eq.~\ref{eq:s-tau} and Eq.~\ref{eq:tau-em} to obtain $r_{\rm calc}$ and EM simultaneously for the 39 sources that are detected at both   6 and 22.4\,GHz. For the 15 compact sources that are only detected at   6\,GHz, we deconvolved the beam size from $r_{\rm obs6}$ to estimate $r_{\rm calc}$, as $r_{\rm calc}^2 + r_{\rm beam}^2 = r_{\rm obs6}^2$, and $r_{\rm beam}^2 = 0.62^{\prime\prime}\times0.28^{\prime\prime}$ for 6\,GHz.
    Then we calculated EM using the estimated $r_{\rm calc}$.
    The $r_{\rm calc}$ and EM of all the compact sources are listed in Table\,\ref{t:coreparams_derived}; those derived from the deconvolved $r_{\rm calc}$ are flagged  with and asterisk (*) in Table\,\ref{t:coreparams_derived}.

    \begin{figure}[ht]
          \begin{center}
          \includegraphics[width=0.37\textwidth]{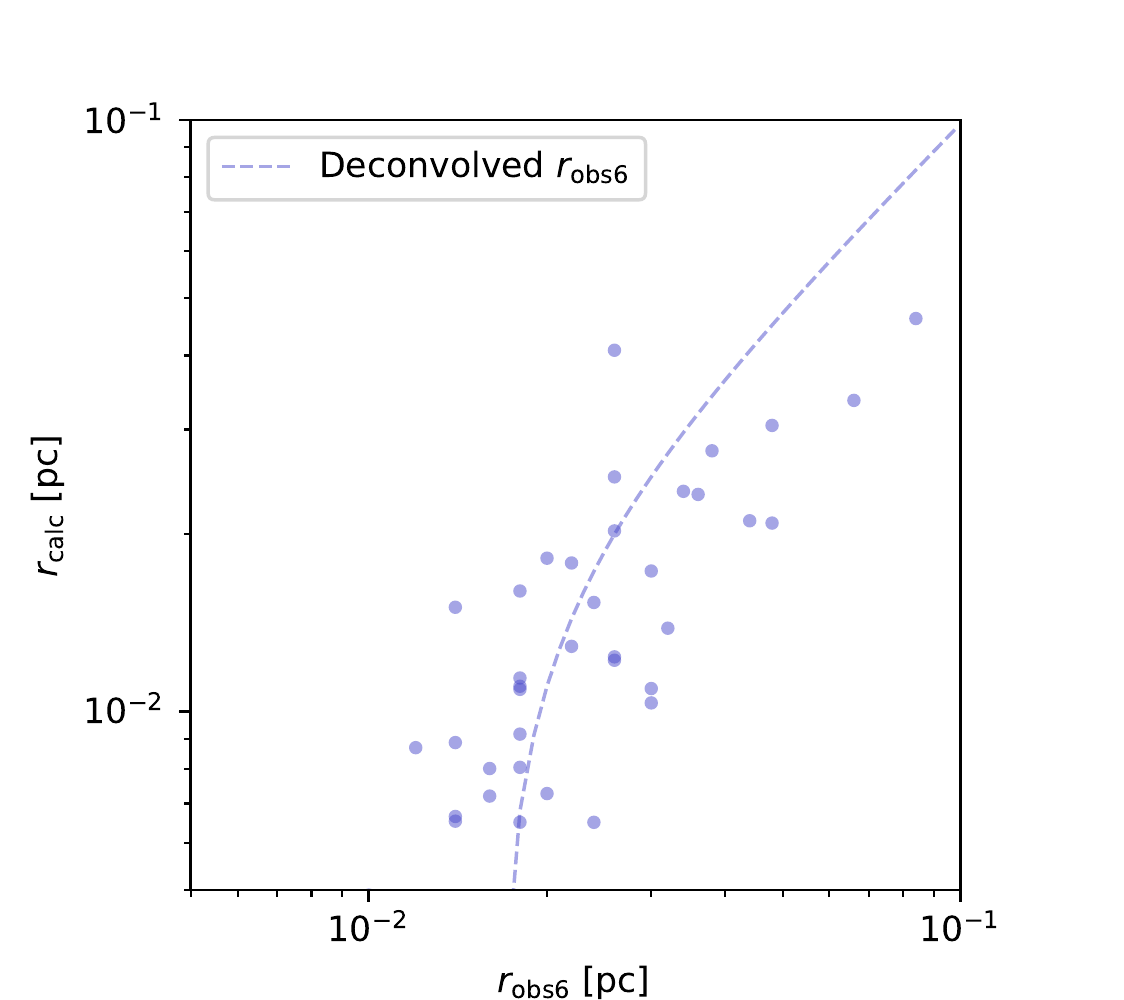}
          \caption[RMS map of 6\,cm image.]{Relation between $r_{\rm obs6}$ and $r_{\rm calc}$ of the 39 compact sources with 22.4\,GHz detection. The dashed curve indicates the deconvolution relationship $r_{\rm calc}^2 + r_{\rm beam}^2 = r_{\rm obs6}^2$, where $r_{\rm beam}$ is the effective radius of the beam. 
           }
          \label{f:r_calc_r_obs}
          \end{center}
    \end{figure}

    \begin{figure}[ht]
            \begin{center}
            \includegraphics[width=0.4\textwidth]{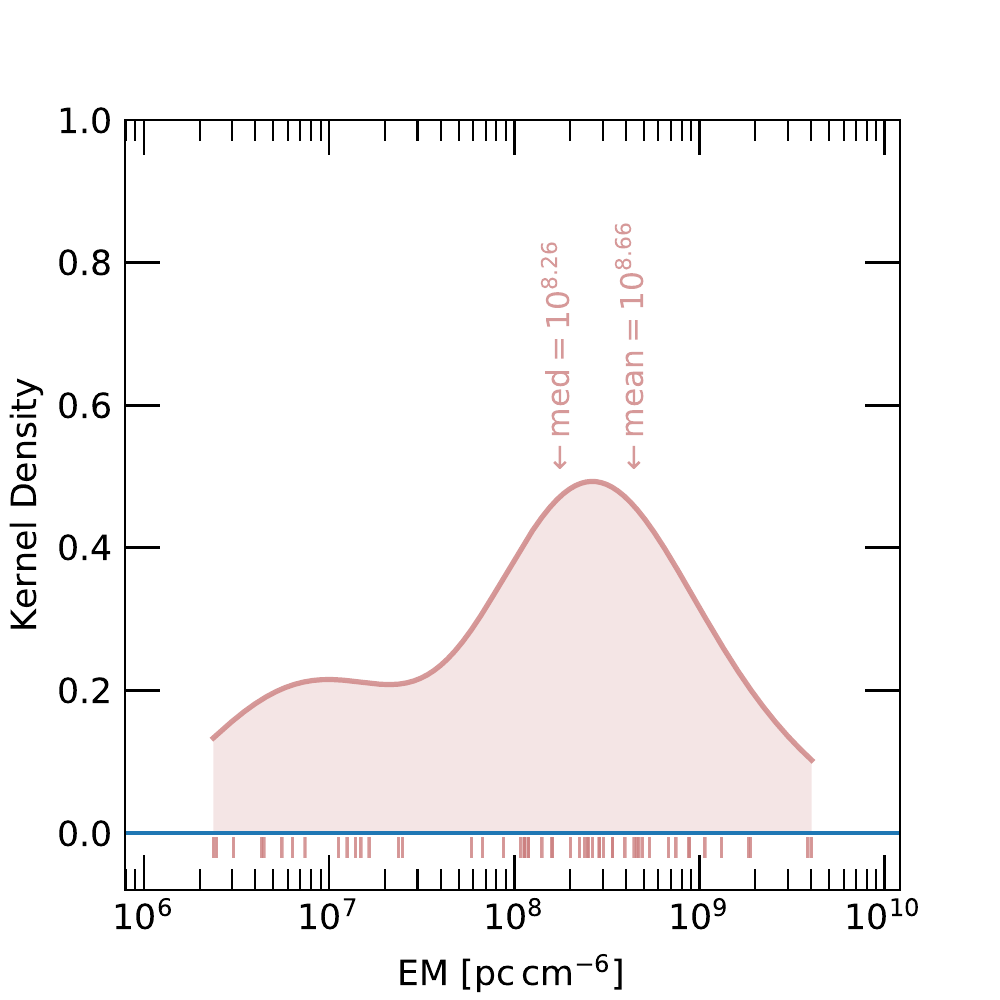}
            \caption[RMS map of 6\,cm image.]{Distribution of EM. The mean and median values are indicated.
             }
            \label{f:em_dist}
            \end{center}
    \end{figure}

    \begin{figure}[ht]
            \begin{center}
            \includegraphics[width=0.4\textwidth]{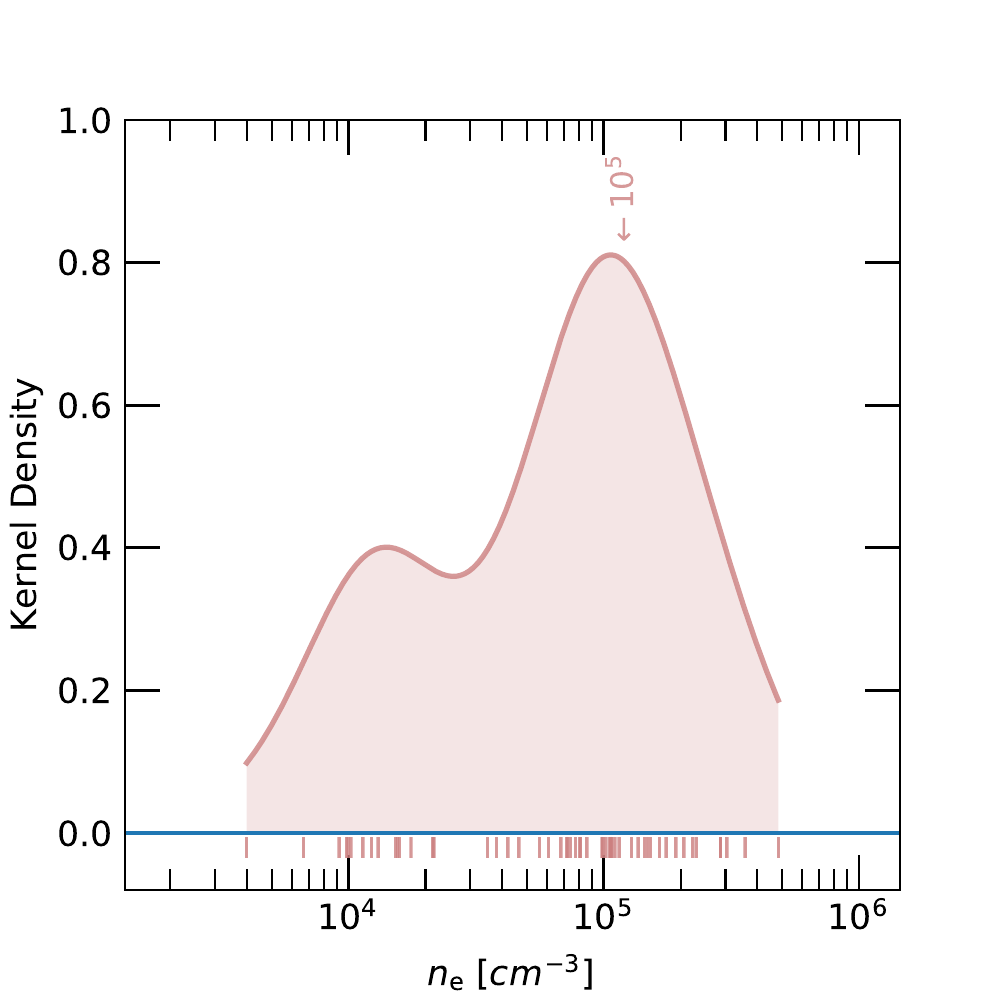}
            \caption[RMS map of 6\,cm image.]{Distribution of $n_{\rm e}$. The mean and median values, which are both $\sim 10^5\,{\rm cm^{-3}}$, are indicated.
             }
            \label{f:ne_dist}
            \end{center}
    \end{figure}

    For all the compact sources that are associated with the 22.4\,GHz detection, we plot the $r_{\rm obs6}-r_{\rm calc}$ diagram (see Fig.\,\ref{f:r_calc_r_obs}).
    We can see that in general $r_{\rm calc}$ follows the trend of deconvolution  (i.e., $r_{\rm calc}^2 =  r_{\rm obs6}^2 - 0.62^{\prime\prime}\times0.28^{\prime\prime}$), which suggests that the method we used to calculate  $r_{\rm calc}$, although without measuring the compact source size in the image, is in general consistent with the size we observed.
    The deviation between the  $r_{\rm calc}$ from the dashed curve in Fig.\,\ref{f:r_calc_r_obs} is likely due to some uncertainties in the observations and flux measurements.
    For example, the determination of $r_{\rm obs6}$ is made by including as much of the flux as possible, which might be larger than the actual observed size.
    In addition, due to the artifacts in the image, we cannot ideally define the boundaries of the compact sources, but only approximate the compact source as a circle with $r = r_{\rm obs}$. 

    \begin{figure}[ht]
          \begin{center}
          \includegraphics[width=0.47\textwidth]{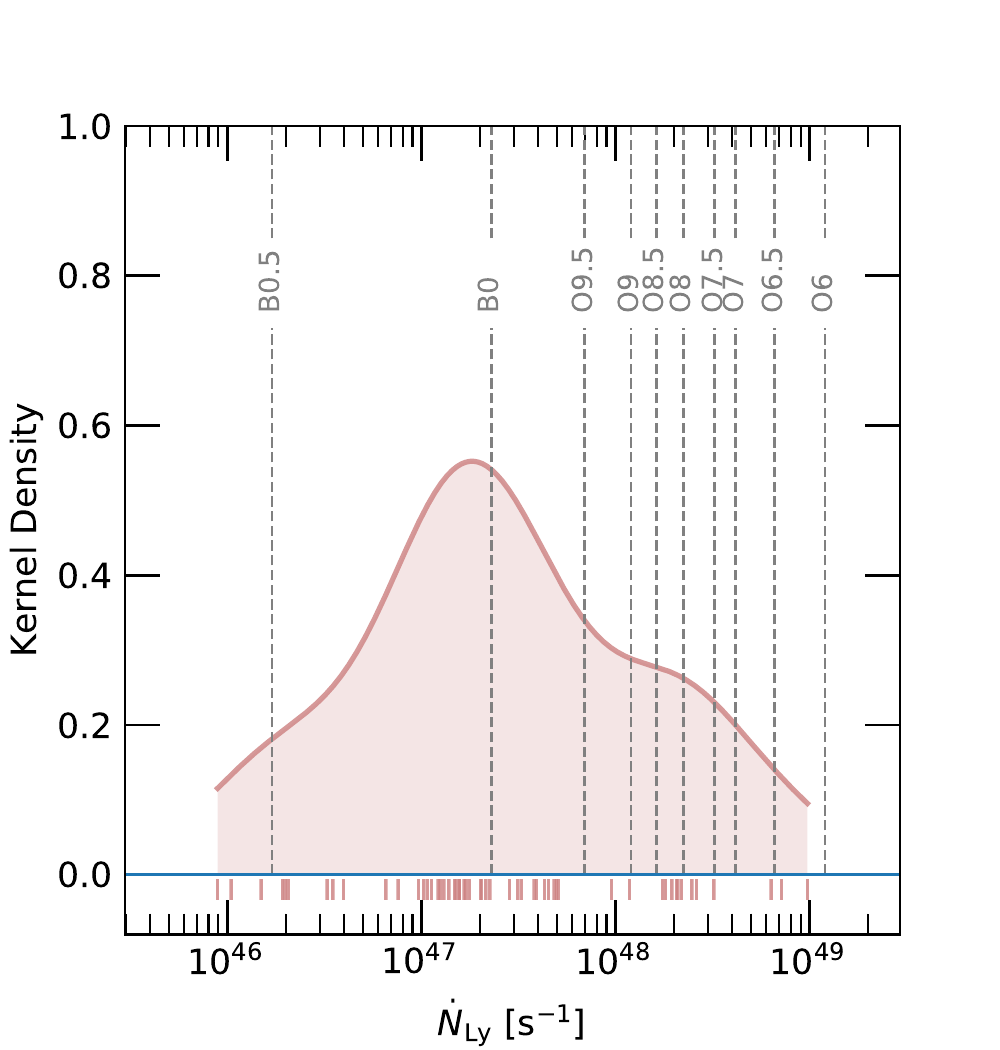}
          \caption[RMS map of 6\,cm image.]{Distribution of $\dot{N}_{\rm Ly}$. The corresponding spectral type of the ionizing stars (assuming that the \hii regions are ionized by single stars) are indicated.
           }
          \label{f:nly_dist}
          \end{center}
    \end{figure}

    The probability distribution of EM of all the 54 compact sources is shown in Fig.\,\ref{f:em_dist}.
    The EM of these compact sources ranges from $\sim 2\times10^6\,{\rm pc\,cm^{-6}}$ to $\sim 3\times10^9\,{\rm pc\,cm^{-6}}$.
    Only eight compact sources have ${\rm EM}<10^7\,{\rm pc\,cm^{-6}}$, which means that  most of the compact sources are consistent with the parameters derived for UC\hii regions \citep[see][]{Kurtz:2002aa}.
    From EM we estimated the electron density $n_{\rm e}$ as $n_{\rm e} = [{\rm EM}/(2r_{\rm calc})]^{1/2}$.
    The distribution of $n_{\rm e}$ is shown in Fig.\,\ref{f:ne_dist}.
    The $n_{\rm e}$ of the 54 \hii regions are from $\sim4\times10^3\,{\rm cm^{-3}}$ to $\sim5\times10^5\,{\rm cm^{-3}}$ with a mean value of $\sim10^5\,{\rm cm^{-3}}$.

    Knowing EM, we calculated the flux of Lyman continuum photons ($\dot{N}_{\rm Ly}$) that are needed to ionize these UC\hii regions following Eq.\,C.18 in \citet{Schmiedeke:2016uc}.
    The derived $\dot{N}_{\rm Ly}$ ranges from $10^{46}\,{\rm s^{-1}}$ to $10^{49}\,{\rm s^{-1}}$ (see Fig.\,\ref{f:nly_dist}).
    If we assume that these cores are ionized by single stars, these UC\hii regions are ionized by stars from spectral types B0.5 to O6 \citep{Panagia:1973aa}.  

\begin{figure*}[ht]
          \begin{center}
          \includegraphics[width=0.33\textwidth]{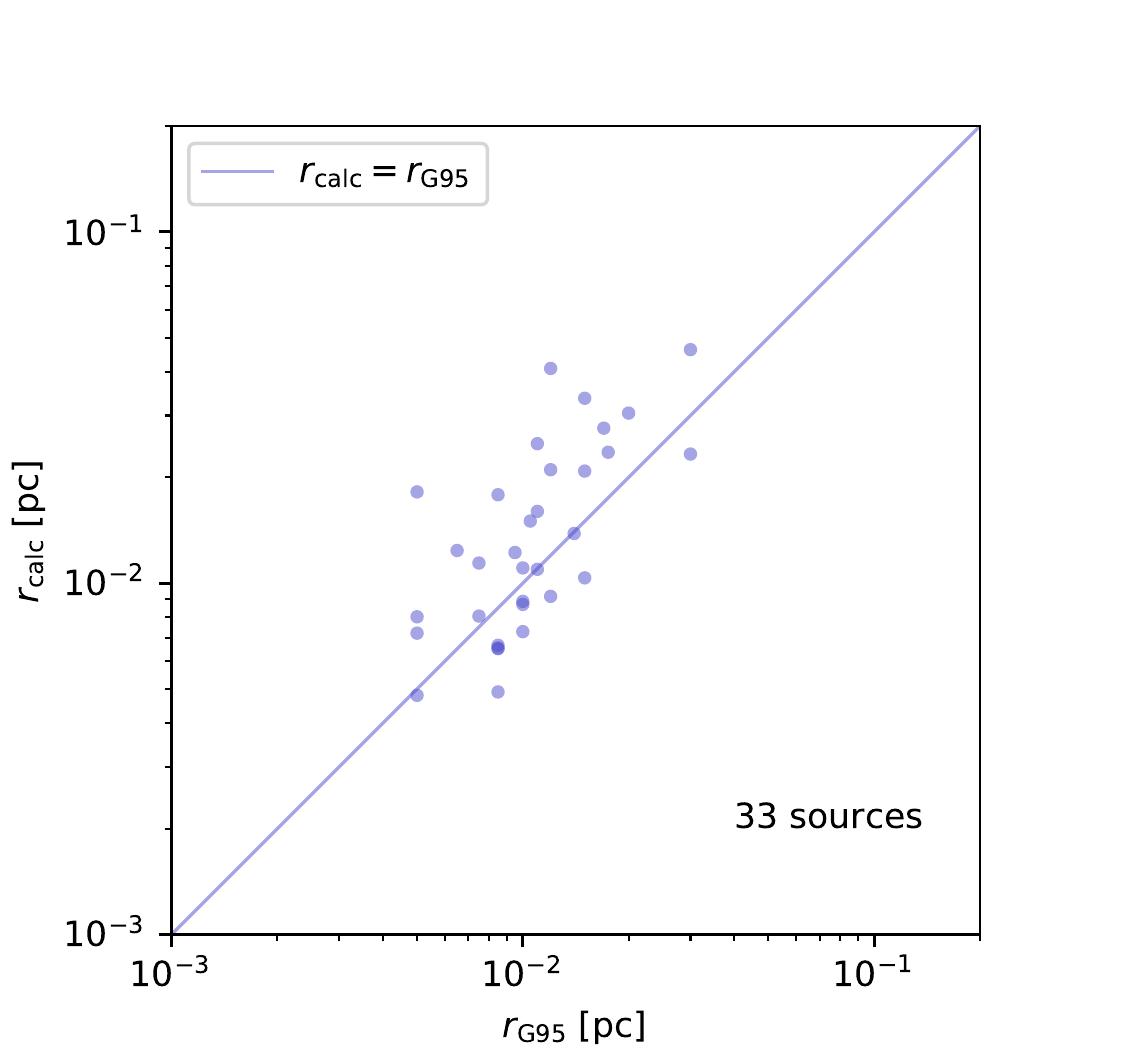}
          \includegraphics[width=0.33\textwidth]{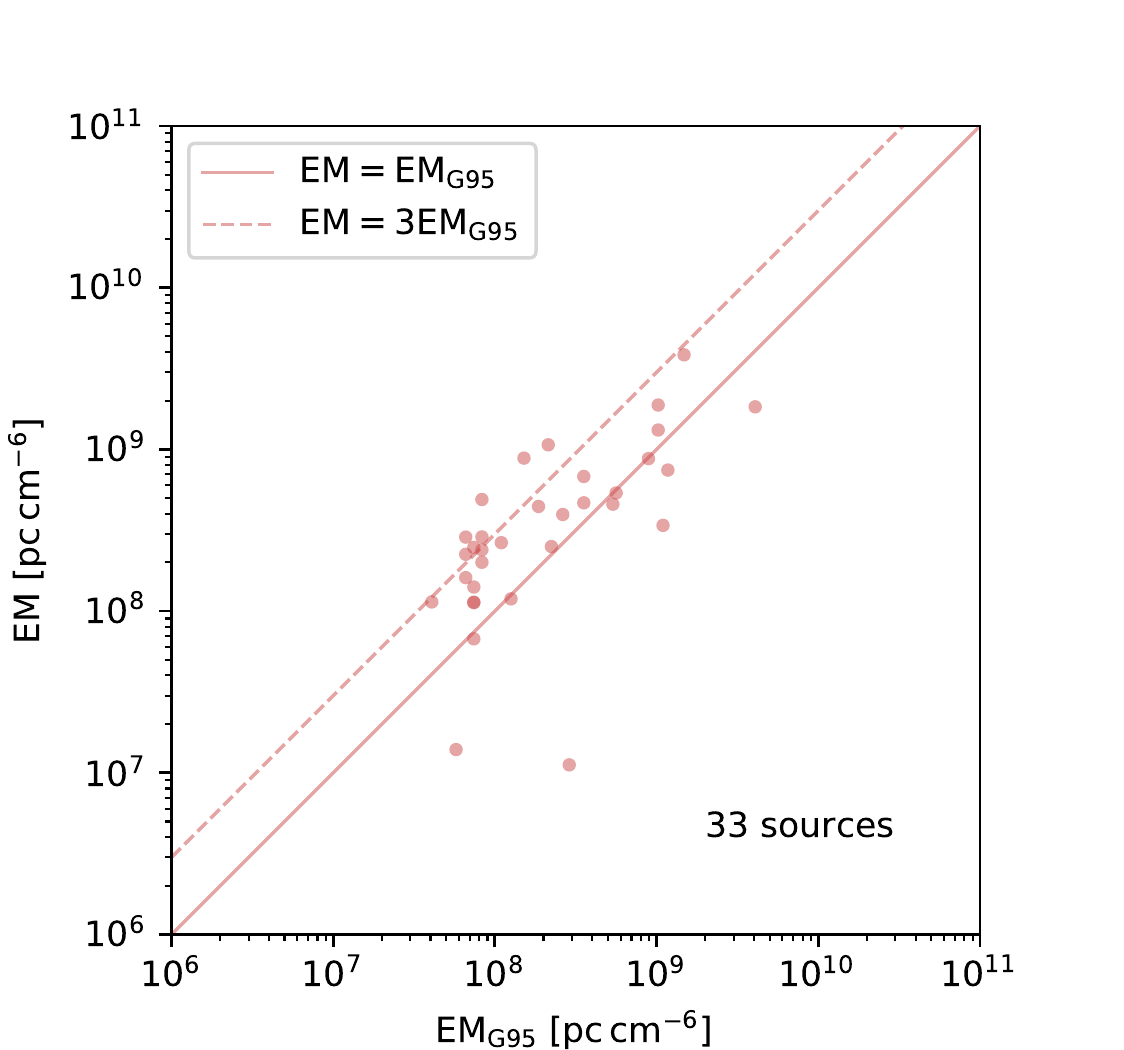}
          \includegraphics[width=0.33\textwidth]{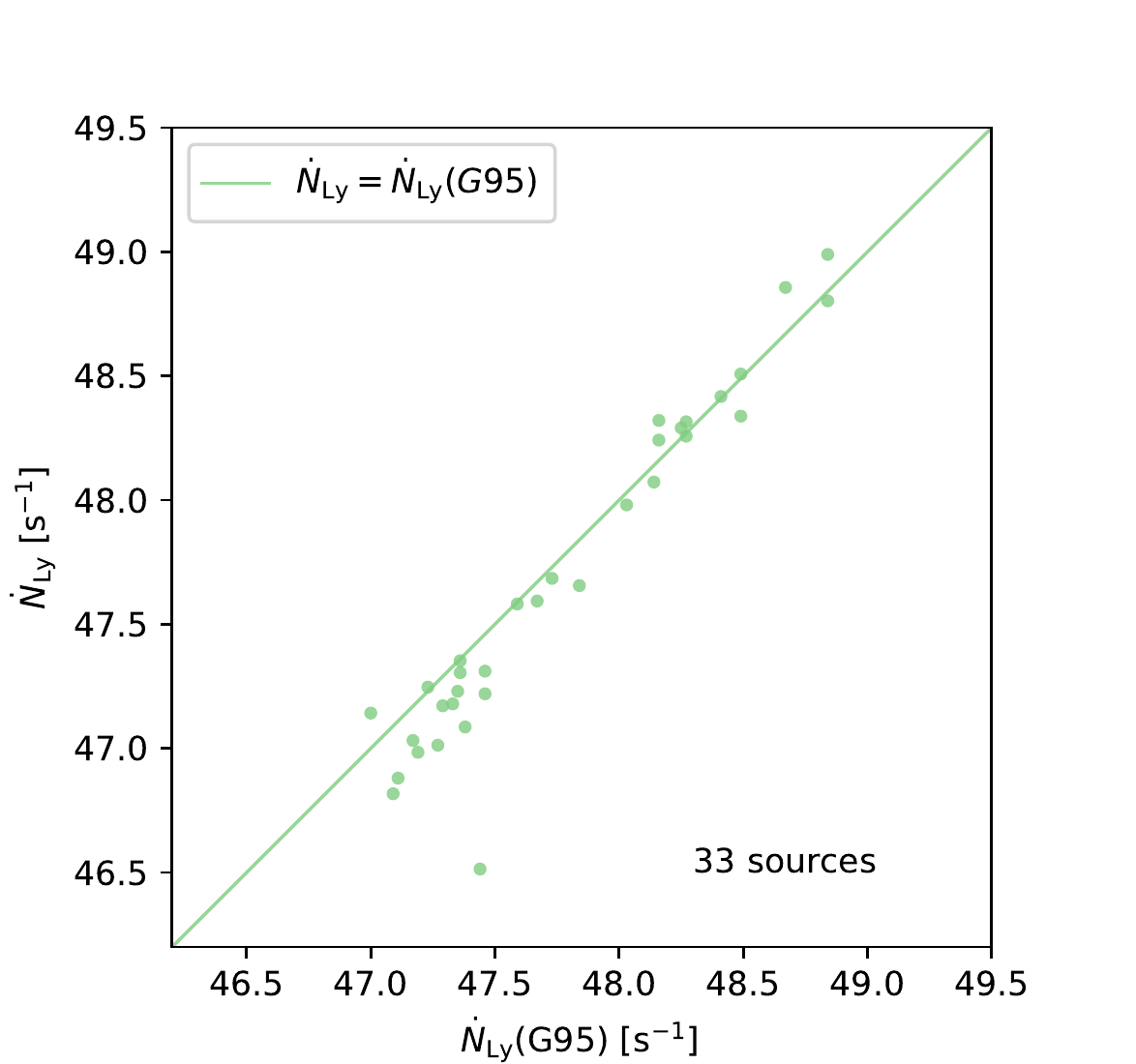}
          \caption[Comparison of R, EM, and nly.]{Comparison between $r_{\rm calc}$ (\emph{left}), EM (\emph{middle}), and $\dot{N}_{\rm Ly}$ (\emph{right}) of 33 sources and the corresponding parameters found by \citet{Gaume:1995aa}.
           }
          \label{f:compare_remnly}
          \end{center}
    \end{figure*}

    In this section we describe a method that is independent of the observed source size to derive the radius ($r_{\rm calc}$) and emission measure (EM) of 39 sources detected at 6 and 22.4\,GHz.
    To verify this a method, we compared the $r_{\rm calc}$, EM, and $\dot{N}_{\rm Ly}$ values of these 39 sources with the values presented by \citet{Gaume:1995aa}, namely $r_{\rm G95}$, $\rm EM_{G95}$, and $\dot{N}_{\rm Ly\,G95}$. 
    Of the 39 sources, 33 are coincident with the UC\hii regions reported by \citet{Gaume:1995aa}.
    Twenty of the 33 sources     have $r_{\rm calc}>r_{\rm G95}$ (see  left panel of Fig.\,\ref{f:compare_remnly}).
    The order of magnitude of EM of the UC\hii regions is consistent in these two studies. Of the 33 sources, 23 have EM larger than $\rm EM_{G95}$ (see   middle panel of Fig.\,\ref{f:compare_remnly}).
    Most of the sources have  $\dot{N}_{\rm Ly}$ similar to $\dot{N}_{\rm Ly\,G95}$ (see   right panel of Fig.\,\ref{f:compare_remnly}), except source 44.
    The difference between $\dot{N}_{\rm Ly}$ and $\dot{N}_{\rm Ly\,G95}$ of source 44 is possibly caused by the time-domain flickering of the radio emission in SgrB2 (see the last paragraph of this section).

    \begin{figure}[ht]
      \begin{center}
          \includegraphics[width=0.35\textwidth]{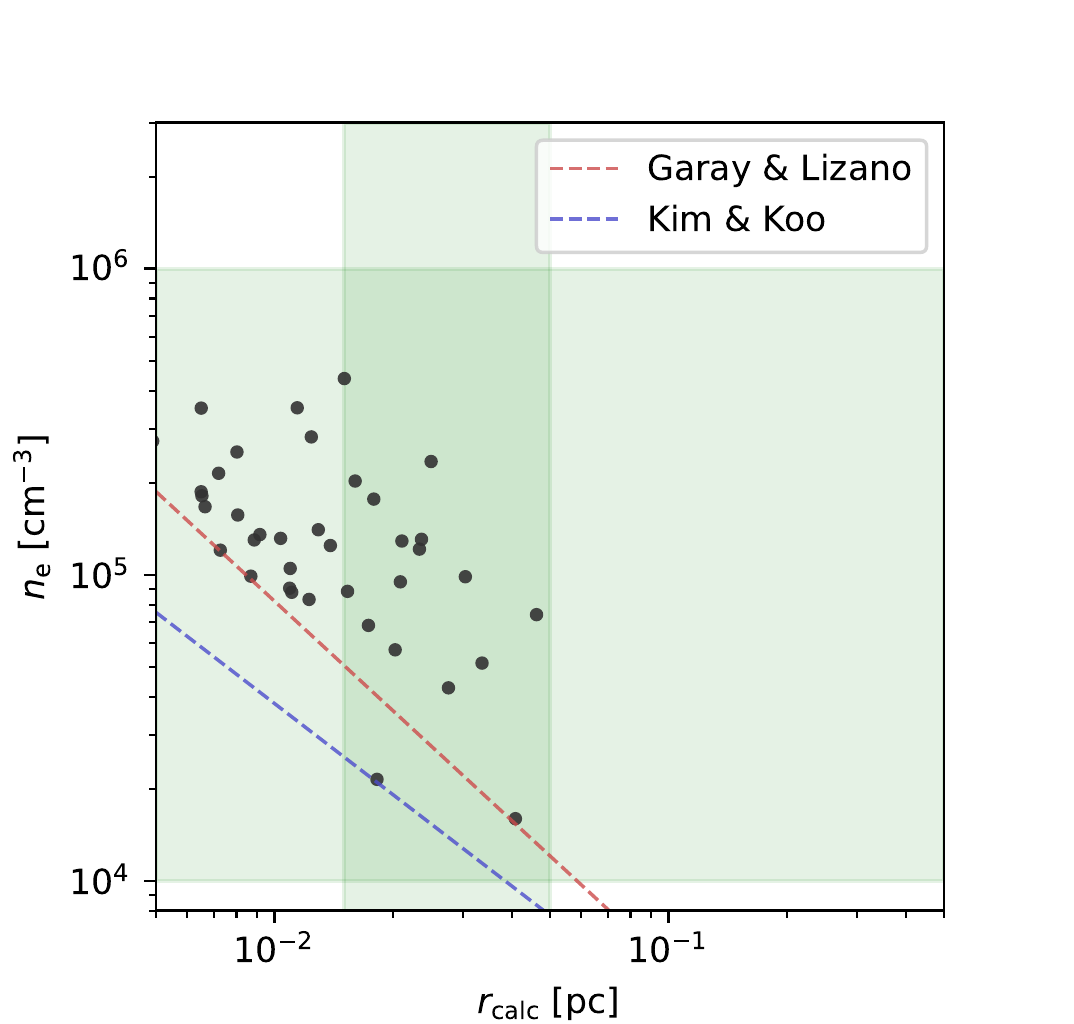}
          \caption[Comparison of R]{Diagram of $r_{\rm calc}$ and $n_{\rm e}$. Previously derived $r_{\rm calc}$--$n_{\rm e}$ relationships for UC\hii regions from \citet{Garay:1999ta}, $n_{\rm e} = 780\times(2r)^{-1.19}$, and from \citet{Kim:2001vy}, $n_{\rm e} = 790\times(2r)^{-0.99}$, are plotted as dashed lines. The green shading indicates the typical $n_{\rm e}$ of UC\hii regions, which is from $\sim 10^4\,{\rm cm^{-3}}$ to $\sim 10^6\,{\rm cm^{-3}}$, and the typical $r$ of UC\hii regions, which is $\sim0.015$ to $\sim 0.05$\,pc \citep{Kurtz:2005aa}.
           }
          \label{f:d_ne}
      \end{center}
    \end{figure}

    In Fig.\,\ref{f:d_ne} we plot the diagram of $r_{\rm calc}$ and $n_{\rm e}$. 
    It is worth noting that all of the 39 \hii regions have typical $n_{\rm e}$ of UC\hii regions (i.e., $\sim 10^4\,{\rm cm^{-3}}$ to $\sim 10^6\,{\rm cm^{-3}}$; \citealt{Kurtz:2005aa}). However, 16 of these have $r$ smaller than that of typical UC\hii regions ($\sim0.015$ to $\sim 0.05$\,pc).
    Although  some of the sources have sizes similar to HC\hii regions, considering the low density of our \hii regions compared to typical HC\hii regions, and following the previous nomenclature of most of the sources \citep[e.g.,][]{Gaume:1995aa}, we still call all the \hii regions in this study  UC\hii regions.

    Recently, \citet{2020ApJ...899...94R} identified similar \hii regions in W\,51 \LEt{ in the W51 complex? }that have $n_{\rm e}\sim10^4-10^5\,{\rm cm^{-3}}$ and  $2\times r\sim 10^{-3}-10^{-2}\,{\rm pc}$. They categorized these sources as HC\hii regions, but they are similar to smaller UC\hii regions and are ionized by early B-type stars, which is in agreement with the spectral type of the ionizing stars for most of our UC\hii regions (see Fig.\,\ref{f:nly_dist}).
    Unlike the HC\hii regions in W51, our UC\hii regions do not exactly follow the $r-n_{\rm e}$ relationships of UC\hii regions proposed by \citet{Garay:1999ta} and \citet{Kim:2001vy}, and have higher $n_{\rm e}$ (by a factor of $\sim 2$) than the predicted values (see Fig.\,\ref{f:d_ne}).
    Such a discrepancy might be due to the neutral gas surrounding our UC\hii regions that is denser ($n_{\rm H_2}\gtrsim 10^6\,{\rm cm^{-3}}$) than typical molecular cores ($n_{\rm H_2}\sim 10^4-10^5\,{\rm cm^{-3}}$; see, e.g., \citealt{Bergin:2007aa}).
    In Sect.\,\ref{sub:dustdensity} a detailed analysis of the gas density surrounding the UC\hii regions is presented. 

    Since we cannot resolve the detailed morphologies of the UC\hii regions, we neglect the possible inhomogeneity of the \hii regions.
    The various morphologies of UC\hii regions result in modified SEDs other than that described by Eq.~\ref{eq:s-tau} and Eq.~\ref{eq:tau-em}, \citep[see, e.g.,][]{Keto:2003ud,Keto:2008wu}.
    Additionally, if there are accretion flows to the \hii regions, flickering of the flux on a  timescale of $\sim 100$\,yr may occur, which is observed by \citet{De-Pree:2014aa} and modeled by \citet{Peters:2010ug,Peters:2010ta}.
    Since the 6\,GHz observation (2013) was performed 23 years after the 22.4\,GHz observations (1989), the effects of flickering could be present in some of these cores. Hence, the simultaneous use of 6\,GHz and 22.4\,GHz fluxes may not be appropriate for the analysis of some sources.
    However, \citet{De-Pree:2014aa} reported the flickering of 4 out of 41 sources in Sgr\,B2 within a similar time range (1989--2014).
    Such a rarity of cases (10\%)  suggests that the flickering may not significantly alter our statistical results.

  \subsection{Dense gas environment} 
  \label{sub:dustdensity}

    We extrapolated the \hii regions' SED (Eq.\,\ref{eq:s-tau}) to subtract the free-free contribution from $S_{96}$ to get the emission purely from dust, which is denoted    $S_{\rm dust}$.
    Twelve UC\hii regions have $S_{\rm dust}$ below $3\times$rms, while the other 42 UC\hii regions have detectable dust emission after subtraction of the free-free emission.
    Using $S_{\rm dust}$, we evaluated the dust properties in the vicinity of the 42 UC\hii regions.

    Following \citet{Ossenkopf:1994aa}, we calculated the dust column density $N_{\rm d}$ as
    \begin{equation}
    \label{eq:dust_sed}
      S_{\rm dust} = \frac{2h\nu^3}{c^2}
      \frac{1}{e^{\frac{h\nu}{k_{\rm B}T_{\rm d}}}-1}
      \left(
      1-e^{
      -\kappa_{0}\left(\frac{\nu}{\nu_0}\right)^{\beta}N_{\rm d}
      }
      \right)
      \frac{\pi r^2}{D^2},
    \end{equation}
    in which $T_{\rm d}$ is the dust temperature, $D$ is the distance of Sgr\,B2, and $r$ is the radius of the dust core.
    Since most of the dust cores are not resolved \citep{Ginsburg:2018wo}, here we use $r = (r_{\rm obs96}^2-r_{\rm beam}^2)^{1/2}$, where $r_{\rm beam} = (0.54^{\prime\prime}\times0.46^{\prime\prime})^{1/2}$ is the effective radius of the 96\,GHz beam.
    The dust parameters $\kappa_{0}$ and $\beta$ depend on the dust grain properties.
    Due to the general high temperature and high density of neutral gas revealed by previous studies \citep[see, e.g.,][]{Huttemeister:1993tr,Schmiedeke:2016uc}, we assume that $T_{\rm d} = 100$, $\kappa_{0} = 2.631$, and $\beta = 1.05$ for $\nu_0 = 100\,{\rm GHz}$.
    The assumption of $\kappa_{0}$ and $\beta$ corresponds to dust grains without ice mantles and with a volume density of $10^8\,{\rm cm^{-3}}$.
    We assume a gas-to-dust mass ratio of 100 \citep{Ott:2014wk,Giannetti:2017vb} to estimate the gaseous mass $N_{\rm H_2}$, $N_{\rm H_2} = 100N_{\rm d}$. 

\begin{figure}[ht]
          \begin{center}
          \includegraphics[width=0.35\textwidth]{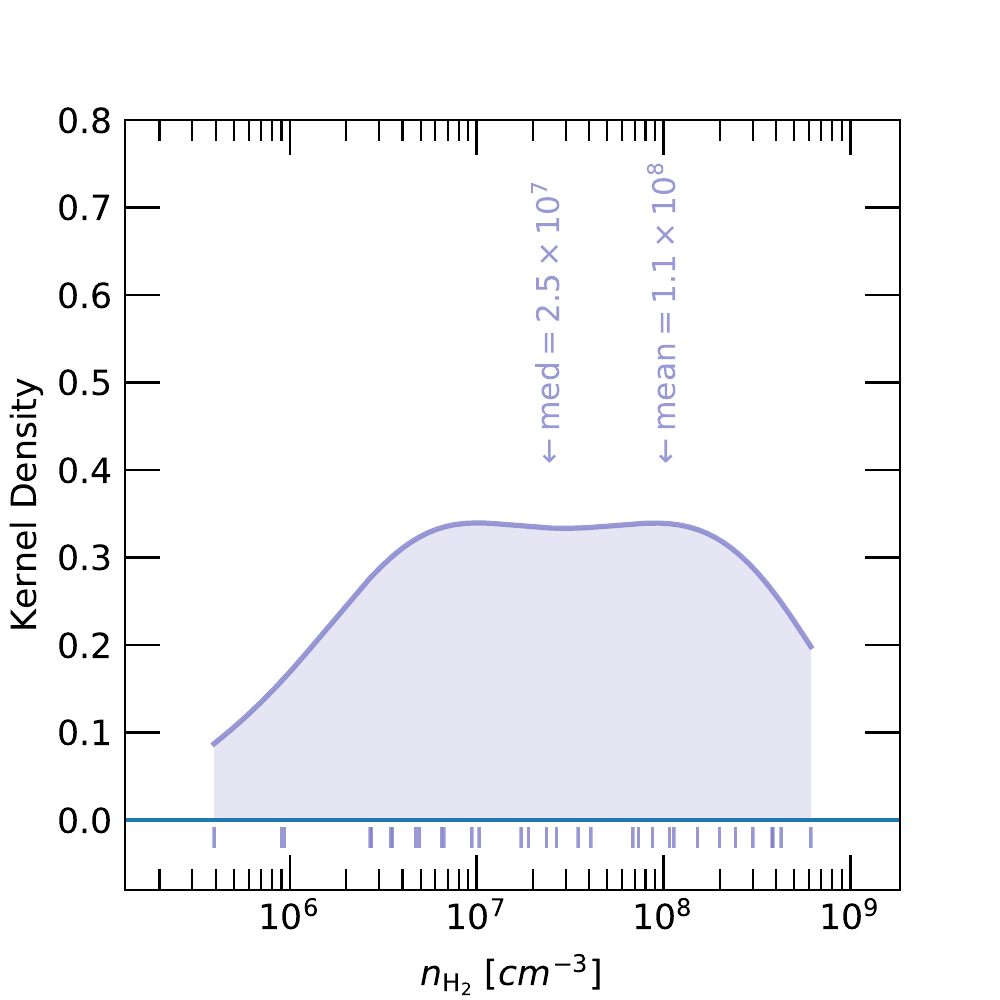}
          \caption[Comparison of R]{Probability distribution of $n_{\rm H_2}$ of the dust   surrounding the UC\hii regions that are associated with dust emission. Median and mean values are indicated.}         
          \label{f:nh2_dist}
          \end{center}
    \end{figure}

    From the column density of molecular gas, $N_{\rm H_2}$, we obtain the volume density of molecular gas (\ce{H2}),
    \begin{equation}
      n_{\rm H_2} = \frac{3N_{\rm H_2}}{4r}.
    \end{equation}
    We list the $n_{\rm H_2}$ values at the position of the 42 UC\hii regions associated with dust emission in Table\,\ref{t:coreparams_derived}.
    The distribution of $n_{\rm H_2}$ is shown in Fig.\,\ref{f:nh2_dist}. The $n_{\rm H_2}$ ranges from $\sim 10^6$ to $\sim 10^8\,{\rm cm^{-3}}$.
    Due to the artifacts and possible small size of the dust core, even a dust core with enough volume density might not be detected by the sensitivity of our data set.
    We used the method in \citet{Ginsburg:2018wo} to estimate the detection limit of the dust emission. 
    Compared to the typical $n_{\rm H_2}$ of the molecular clouds surrounding UC\hii regions ($10^5\,{\rm cm^{-3}}$) \citep[e.g.,][]{Wood:1989aa}, our UC\hii regions reside in denser neutral gas.
    The median value obtained in this work is similar to the value ($2\times10^7\,{\rm cm^{-3}}$) measured in the central region of  Sgr B2(M) \citep{de-Pree:1995ve,Huettemeister:1995aa,Sanchez-Monge:2017vt}, although   our sources are also distributed   in the outskirts of Sgr B2(M).

\section{Analysis and discussion} 
\label{sec:analysis_and_discussion}

  \subsection{Expansion and equilibrium} 
  \label{sub:expansion_time}

      \begin{figure*}[ht]
      \begin{center}
        \includegraphics[width=0.85\textwidth]{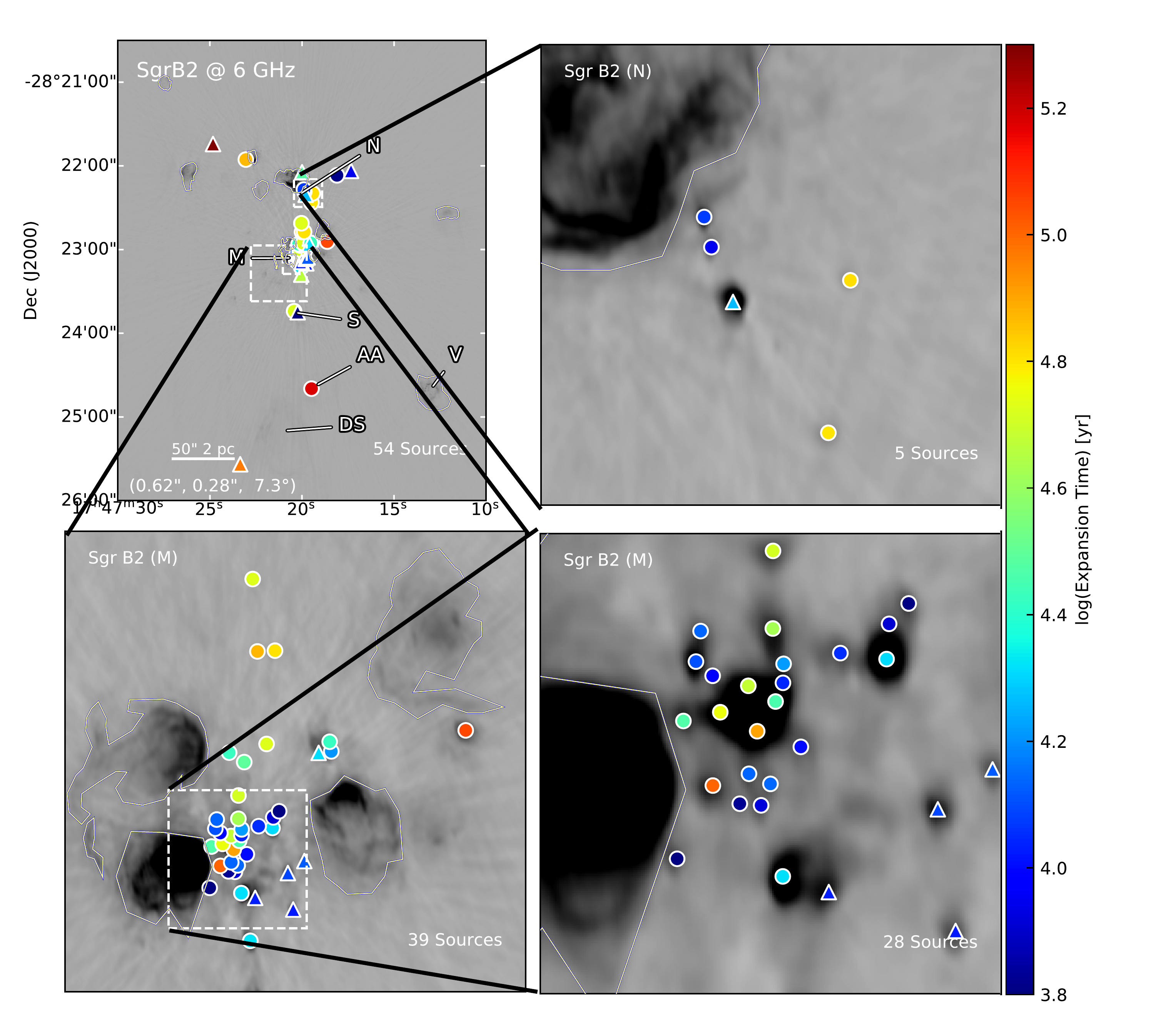}
        \caption[Time distribution]{
        Evolution sequence of the \hii regions. The circles are calculated with the  dust density derived from the  96\,GHz image. The triangles are calculated with the dust density of $2\times10^7\,{\rm cm^{-3}}$. The color of the markers indicates the expansion timescale.
         }
        \label{f:cores_time}
      \end{center}
    \end{figure*}

\begin{figure}[ht]
      \begin{center}
        \includegraphics[width=0.45\textwidth]{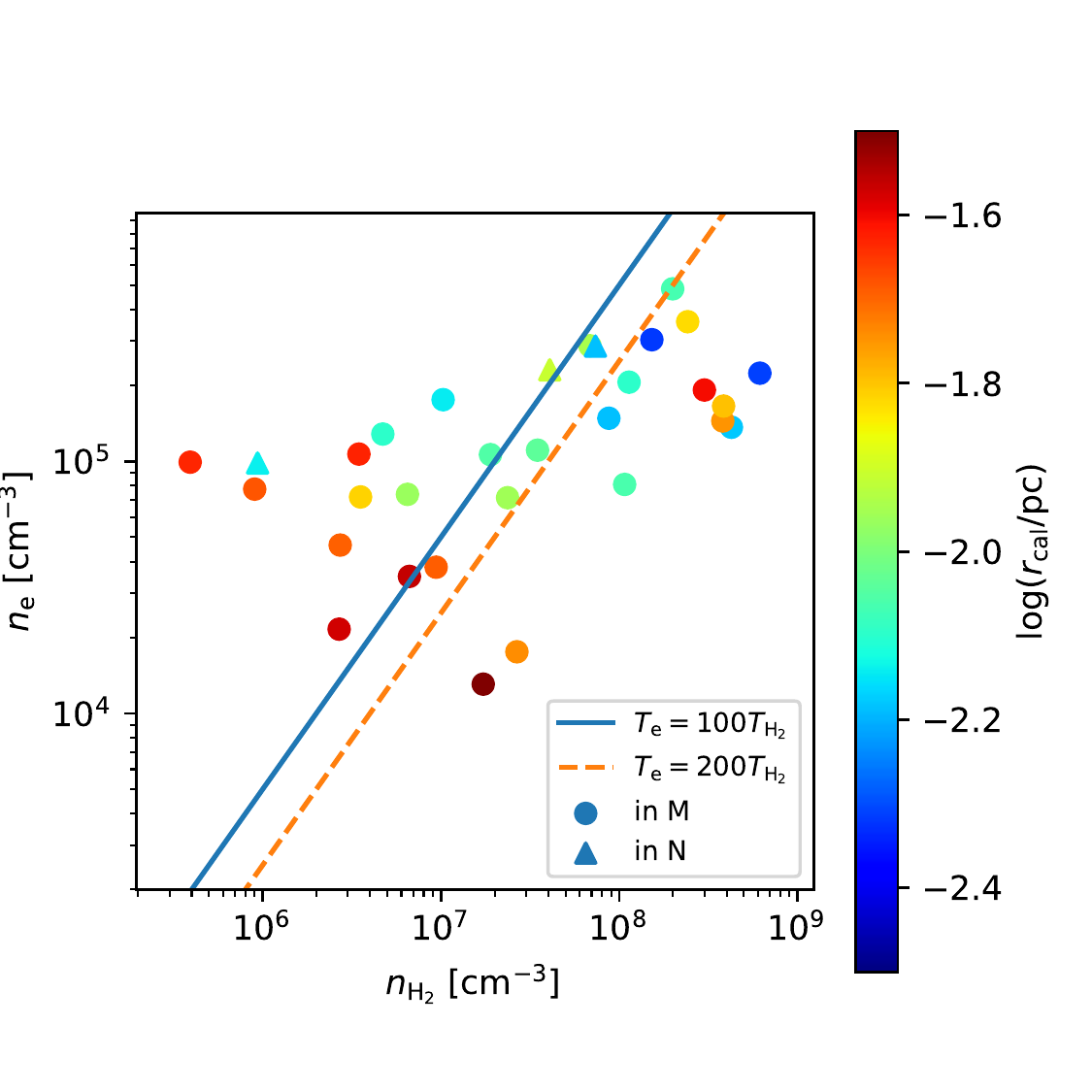}
        \caption[Comparison of R]{
        Diagram of $n_{\rm e}$ and $n_{\rm H_2}$. The \hii regions in Sgr\,B2(M) are shown as circles, while those in Sgr\,B2(N) are shown as triangles. The color indicates the size $r_{\rm calc}$ (see color scale at right). Equilibrium conditions are plotted as solid and dashed lines for $T_{\rm H_2} = 100\,{\rm K}$ and $50\,{\rm K}$ and thus for $T_{\rm e} = 100T_{\rm H_2}$ and $200T_{\rm H_2}$.}
        \label{f:pressure_eq}
      \end{center}
    \end{figure}

    Based on the physical properties of the \hii regions (see Sect.\,\ref{sub:cores_in_vla}) and the properties of their surrounding environment (see Sect.\,\ref{sub:dustdensity}), we now investigate the expansion time for these extremely compact and dense \hii regions.
    The density of the molecular cloud that UC\hii regions expand into significantly affects the expansion rate \citep[e.g.,][]{Wood:1989aa,de-Pree:1995ve,De-Pree:1998aa}.
    For the simple case of a spherical UC\hii region ionized by the  Lyman continuum flux of $\dot{N}_{\rm Ly}$ and expanding in molecular gas with volume density ${n}_{\rm H_2}$ and electron temperature of $10^4\,{\rm K}$, following \citet{de-Pree:1995ve,De-Pree:1998aa} we calculate the initial Str\"{o}mgren radius $r_{\rm i}$ as
    \begin{equation}
      \frac{r_{\rm i}}{\rm pc} = 1.99\times10^{-2}
      \left(\frac{\dot{N}_{\rm Ly}}{\rm 10^{49}\,s^{-1}}\right)^{1/3}
      \left(\frac{{n}_{\rm H_2}}{\rm 10^{5}\,cm^{-3}}\right)^{-2/3}.
    \end{equation}
    For the 12 UC\hii regions without detected associated dust emission ($S_{\rm dust}$, which is $S_{96}$ subtracting the free-free component), we assumed a uniform dust density of $2\times10^7\,{\rm cm^{-3}}$ following \citet{De-Pree:2015ve}.
    This value is in agreement with the median of $n_{\rm H_2}$ of the 42 sources that have physical $S_{\rm dust}$.
    Since   we assumed  a relatively high gas density compared to typical dust cores \citep[see, e.g.,][]{Bergin:2007aa}, the nondetection of $S_{\rm dust}$ for 12 sources under such an assumption is due to small $r$ but not low $n_{\rm H_2}$.
    Then we calculated the expansion timescales of all   54 cores applying the expansion equation by \cite{Spitzer:1968wl,Dyson:1980tp},
    \begin{equation}{}
    \label{eq:uchiiexpansion}
      r_{\rm calc} = r_{\rm i}
      \left(1 + \frac{7c_{\rm i}t}{4r_{\rm i}}\right)^{4/7},
    \end{equation}
    where $c_{\rm i}$ is the sound speed ($\sim 10\, {\rm km\,s^{-1}}$).  The expansion times, $t$, of the UC\hii regions are listed in Table\,\ref{t:coreparams_derived}.
    In Fig.\,\ref{f:cores_time}   the cores are color-coded according  to their $t$. In ideal cases most of the cores have an expansion time between $10^4$\,yr and $10^5$\,yr.  The expansion times  in this section were dervied  under the assumption of ideal conditions \citep{Spitzer:1968wl}. Such derived expansion times should be treated as lower limits and the actual expansion time might be longer for to at least three reasons. First. the expansion of UC\hii regions can be halted due to the pressure equilibrium between the ionized gas and the surrounding dense molecular gas \citep[e.g.,][]{De-Pree:1998aa}.
    We assume that $T_{\rm e}$ of the ionized gas is $10^4\,{\rm K}$, and the molecular temperature $T_{\rm H_2}$ is from $50\,{\rm K}$ to $100\,{\rm K}$. 
    The pressure equilibrium condition ($2n_{\rm e}T_{\rm e}=n_{\rm H_2}T_{\rm H_2}$) can be expressed as $2n_{\rm e} = 5\times10^{-3}n_{\rm H_2}$ and $2n_{\rm e} = 10^{-2}n_{\rm H_2}$ for $T_{\rm H_2} = 50\,{\rm K}$ and $100\,{\rm K}$, respectively.
    In dust cores with $n_{\rm H_2} \gtrsim \times10^7\,{\rm cm^{-3}}$, the UC\hii regions are mostly in equilibrium with the neutral gas, evident from Fig.\,\ref{f:pressure_eq}, whereas for the  lower $n_{\rm H_2}$ regime, the UC\hii regions are in expansion phase owing to ionized gas pressure exceeding the neutral gas pressure.  
    Second, in reality, dust absorption may reduce $r_{\rm i}$, and therefore slows down the expansion \citep[see, e.g.,][]{Wood:1989aa,De-Pree:1998aa}.
    Third, accretion flow on to the central star will disrupt the expansion process and cause a sudden decrease in the flux and size of the UC\hii region. \citep[see, e.g., Fig.\,10 in][]{Peters:2010tv}. With the current data, we cannot quantify the effect of either of these mechanisms.
    Future observations, for example  high-resolution radio recombination line observations, may provide constraints on the possible effect of accretion on the expansion of the \hii regions.

  \subsection{Evolutionary stages} 
  \label{sub:evolutionary_stages}

    \begin{table}
    \caption[Associated objects of the dust cores.]{\meng{Number of objects associated with dust cores}\LEt{ Number of objects associated with the dust cores}}
    \label{t:small.assoc}
    \begin{tabular}{l r r r r r r}
      \toprule
      \hline \noalign{\smallskip}
      Region   & N & M  & S  & DS & Rest$^{\rm a}$ & Total \\
      \hline \noalign{\smallskip}
      Dust core        & 24  & 55  & 46  & 46  & 100  & 271  \\
      \hii region      & 8   & 40  & 2   & 1   & 3   & 54  \\
      Outflow          & --$^{\rm b}$  & --$^{\rm b}$  & --$^{\rm b}$  & 31  & 18$^{\rm b}$& 49$^{\rm b}$\\
      \ce{CH3OH} Maser & 2   & 2   & 0   & 0   & 5    & 9   \\
      \bottomrule           
    \end{tabular}
    {\\\smallskip\textbf{a}: Rest of the envelope.\\
    \textbf{b}: Not fully covered by the \ce{SiO} map.}
    \end{table}

    The type of objects that are associated with dust cores suggests the evolutionary stages of the dust cores in star formation activity.  \meng{In Table~\ref{t:small.assoc} we summarize the type and number of objects associated with the dust cores in the subregions Sgr\,B2(N), Sgr\,B2(M), Sgr\,B2(S), and  Sgr\,B2(DS), and in the rest  of the envelope of Sgr\,B2.}


    Of all the 54 UC\hii regions, 8 are in Sgr\,B2(N), 40 are in Sgr\,B2(M), 2 are in Sgr\,B2(S), 1 is in Sgr\,B2(DS) and 3 are in the rest of the envelope. 
    Although   the dust cores are distributed all over Sgr\,B2 \citep{Ginsburg:2018wo} and more than 80 of them are associated with the large \hii region in Sgr\,B2(DS) \citep{Meng:2019aa}, the dust cores are rarely associated with any UC\hii regions outside of Sgr\,B2(N) and Sgr\,B2(M).
    The percentage of the dust cores that are associated with UC\hii regions are 33\%, 73\%, 4\%, and 1\% for Sgr\,B2(N), Sgr\,B2(M), Sgr\,B2(S), and Sgr\,B2(DS), respectively, while for the remaining part of the envelope the percentage is 3\%.

    For a dust core,  association with the  \hii region is a sign of that the core is  more evolved \citep[see, e.g.,][]{Breen:2010aa}.
    Therefore, the cores in Sgr\,B2(M) are the most evolved, while the cores in Sgr\,B2(S) and Sgr\,B2(DS) are the least evolved.
    The evolutionary stages of the cores in Sgr\,B2(N) are between those of Sgr\,B2(M) and Sgr\,B2(DS) or Sgr\,B2(S). 

\begin{figure}[ht]
      \begin{center}
        \includegraphics[width=0.45\textwidth]{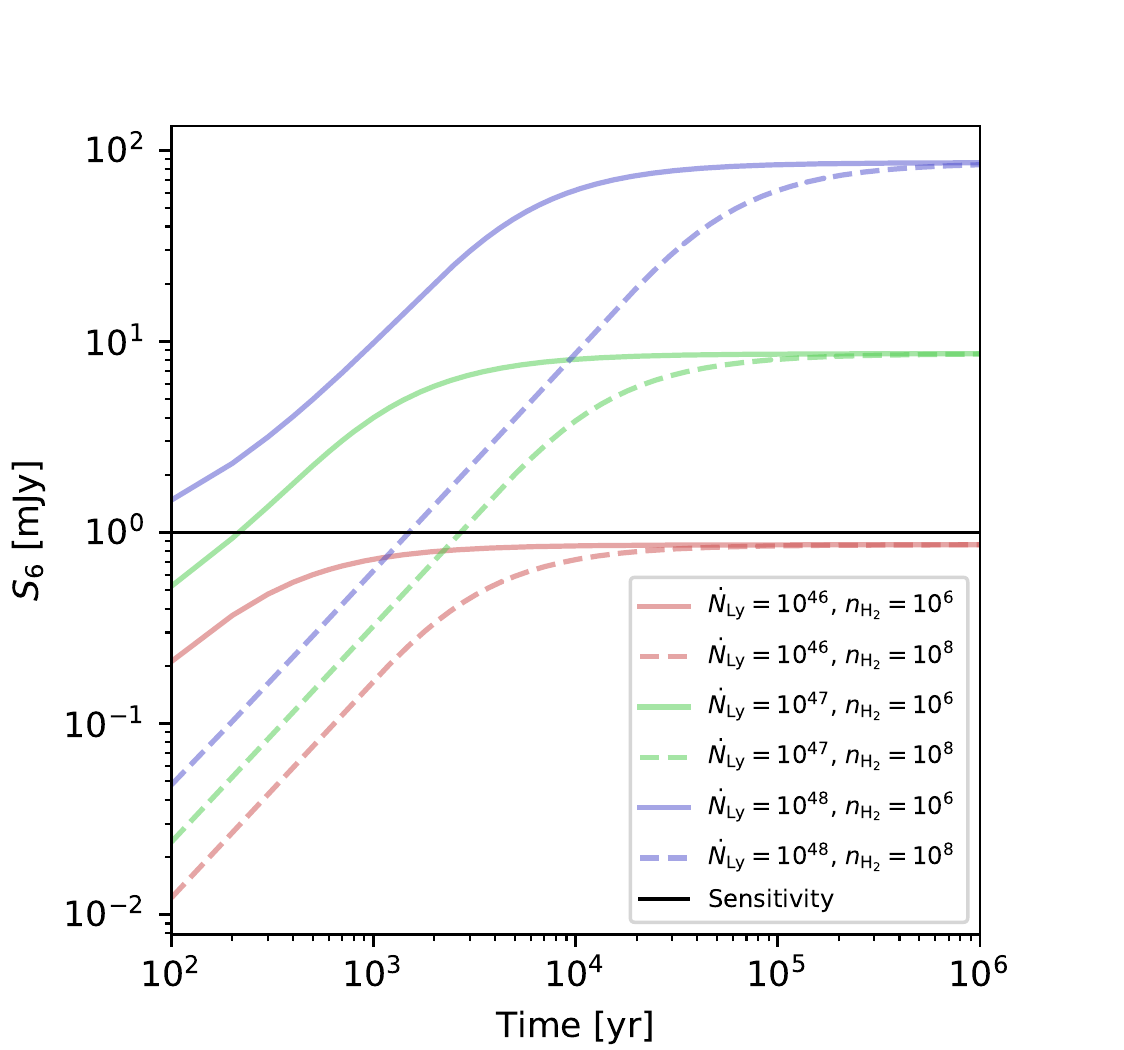}
        \caption[Evolution]{Modeled evolution of $S_6$ of \hii regions with time. The black line denotes the sensitivity of our 6\,GHz observation, which is $\sim 1$\,mJy, for unresolved sources.}
        \label{f:evolution}
      \end{center}
    \end{figure}

    All the dust cores in Sgr\,B2(DS) appear to be pure dust\footnote{
    One \hii region is found associated with a core reported by \citet{Ginsburg:2018wo},  but the core emission is pure free-free, which suggests that the core is a dust-free \hii region. Therefore, no dust cores in Sgr\,B2(DS) are associated with \hii regions.}. 
    With the sensitivity of our observation ($\sim 1$\,mJy, for unresolved sources), we constrain the age of the possibly undetectable embedded \hii regions in the dust cores.
    We consider the $\dot{N}_{\rm Ly}$ from the central star are of $10^{46}$, $10^{47}$, $10^{48}$\,${\rm s^{-1}}$, and dust densities are $10^6$ and $10^8$\,${\rm cm^{-3}}$, which are based on the typical values of the detected UC\hii regions.
    The evolution of $S_6$ of the \hii regions, based on the expansion model
     in Sect.\,\ref{sub:expansion_time} 
     and Eq.~8 in \citep{De-Pree:1998aa}, are displayed in Fig.\,\ref{f:evolution}. 
     The temporal evolution of $r$ and EM leads to increase in $S_{\rm 6}$ (see Eq.\,\ref{eq:s-tau}).
    For the dust cores that have $n_{\rm H_2} = 10^6$\,${\rm cm^{-3}}$, the embedded \hii regions with $\dot{N}_{\rm Ly} \geq 10^{47}$\,${\rm s^{-1}}$ will be observable   $\sim100$\,yr after their birth, which contradicts the nondetection of \hii regions in Sgr\,B2(DS).
    For the dust cores that have $n_{\rm H_2} = 10^8$\,${\rm cm^{-3}}$, the embedded \hii regions with $\dot{N}_{\rm Ly} \geq 10^{47}$\,${\rm s^{-1}}$ will be observable   $\sim10^3$\,yr after their birth, which suggests that it is possible that the \hii regions are already formed but still too dim to be detected.
    The \hii regions with $\dot{N}_{\rm Ly} \leq 10^{46}$\,${\rm s^{-1}}$ will be always undetectable with the current sensitivity. 
    The sizes of the embedded \hii regions will reach the typical size of the UC\hii regions in this study ($\sim 0.01\,{\rm pc}$) in $10^4$ years \citep[see Fig.\,1 in]{de-Pree:1995ve}.
    Thus, it is possible that stars later than B1 have  already formed and ionized the interiors of the dust cores. 

    Collimated outflows are the footprints of the very early stages of star formation activity \citep[see, e.g.,][]{Beuther:2002aa}.
    Outflows can be traced by SiO emission \citep[e.g.,][]{Schilke:1997aa}.
    In a recent project, the SiO~(5--4) emission was observed with ALMA (P.I. A. Ginsburg).
    For the details of the observation and data reduction, see Jeff et. al. (in prep.).
    The resolution is 0.35\arcsec\ $\times$ 0.24\arcsec, with P.A. of $-80^{\circ}$.
    The spectral resolution is 1.35~km~s$^{-1}$. The typical RMS of the image is 0.9~mJy/beam.
    The observation covers Sgr\,B2(S) and the eastern part of Sgr\,B2(DS).
    The peak intensity map of SiO~(5--4) is shown in Figure~\ref{f:small.sio_peak}.
    Due to the artifacts around Sgr\,B2(S), we only analyzed the part of the image with declination $<-$28:24:00.
    We  generated the moment one map by masking out all the pixels below $3\sigma$ (see Fig.~\ref{f:small.sio_mom1}).
    The average velocity difference between the blueshifted and redshifted lobes is found to be $\sim 10$\kms.
    We visually matched the positions of the dust cores   identified by \citet{Ginsburg:2018wo} and the SiO outflows. The dust cores that are 1) spatially associated with an outflow, 2) covered by the SiO image not associated with an outflow, and 3) not covered by the SiO image are indicated  in Table~\ref{t:g18_outflow}.
    Spatially, most of the outflows are associated with dust cores.
    On the other hand, of all the 120 dust cores covered by the SiO image, 49 are identified as associated with outflows. 
    Particularly, among all the cores in Sgr\,B2(DS), two-thirds are associated with outflows.
    Such a high fraction confirms that the cores in Sgr\,B2(DS) are at their very early evolutionary stages.

    \ce{CH3OH} masers serve as   additional probes to trace the star formation activity \citep[see, e.g.,][]{Breen:2010aa}.
    We cross-matched the \ce{CH3OH} masers presented by \citet{Caswell:2010aa} with our dust cores\footnote{We searched for associated masers in a circle with a radius of 1 arcsec, which is double the beam size of the 96\,GHz image.}.
    In Table~\ref{t:small.assoc} the number of masers that are associated with the dust cores that are in each region are listed.
     The cores in   Sgr\,B2(M) and in Sgr\,B2(N) are associated with two \ce{CH3OH} masers.
    The cores in Sgr\,B2(DS) have no associated masers.
    This lack of associated masers also suggests that the cores in Sgr\,B2(DS) are less evolved than those in Sgr\,B2(N) and Sgr\,B2(M).

    It is worth noting that the dust cores in Sgr\,B2(DS) are distributed around the large-scale \hii region in Sgr\,B2(DS) \citep[See Fig.\,2 in][]{Meng:2019aa}.
    The large-scale \hii region was ionized by a central O7 star and is still in the  expansion phase \citep{Meng:2019aa}.
    The expansion time of this \hii region is $\sim 10^5\,{\rm yr}$, which is estimated based on the size of the \hii region (0.72\,pc) and the sound speed of the ionized gas, $10\,{\rm km\,s^{-1}}$.
    This timescale, compared to the expansion time of the \hii regions in Sgr\,B2(M) and Sgr\,B2(N) suggest that the central O7 star   possibly formed before the stars in Sgr\,B2(M) and Sgr\,B2(N).
     Another possible scenario is that the O7 star was ejected from a nearby star forming site (e.g., Sgr\,B1) earlier than $10^5\,{\rm yr}$ ago, and Sgr\,B2 had no star forming activity at that time. The newly formed dust cores in Sgr\,B2(DS), on the other hand, are signs of star forming activity that possibly was triggered by the expansion of the large-scale \hii region in Sgr\,B2(DS).


\section{Summary} 
\LEt{ please include a few lines before and after the bullet list (which acts as a "frame" for the list. A list should never be left "hanging" alone). Something like this: In this  paper we wanted to show ... Our main conclusions can be summarized as follows:   }
\label{sec:summary}

\meng{We  observed the  Sgr\,B2 complex with VLA A, BnC, and D array at 6\,GHz and identified 54 compact radios sources.
We found that 39 of the 54 sources are also detected  in the  22.4\,GHz band \citep{Gaume:1995aa}.
Our main results are summarized as follows:}
\begin{itemize}
  \item 
  Using the 6\,GHz data of all  54 sources, as well as the 22.4\,GHz data of 39 sources, we calculated the EM, radius, electron densities, and the spectral type of the ionizing stars of all the 54 UC\hii regions.
  The UC\hii regions have radius between $6\times10^{-3}\,{\rm pc}$ and $4\times10^{-2}\,{\rm pc}$, and have EM between $10^{6}\,{\rm pc\,cm^{-6}}$ and $10^{9}\,{\rm pc\,cm^{-6}}$.
  We found that the electron densities of these UC\hii regions are in agreement with the values of typical UC\hii regions, while the radii are smaller than for the typical UC\hii regions.
  We identified that the UC\hii regions are ionized by stars with spectral types ranging between B0.5 to O6.

  \item Using the 96\,GHz ALMA data, we characterized the dense gas environment where the UC\hii regions are located.
  We found a typical dense gas density of $\sim10^6-10^9\,{\rm cm^{-3}}$ around the UC\hii regions.
  Using \citet{Spitzer:1968wl}, we estimated the expansion timescale of the UC\hii regions as $\sim10^4-10^5\,{\rm yr}$.
  More than half of the UC\hii regions are close to equilibrium with the neutral gas; this means that  the pressure of most UC\hii regions and the dense gas surrounding them are comparable. Due to the high pressure of the neutral gas, some natal \hii regions might be optically thick and cannot be detected with the sensitivity of this study.
  Instead,  for the  lower $n_{\rm H_2}$ regime ($n_{\rm H_2} \lesssim \times10^7\,{\rm cm^{-3}}$) the ionized gas pressure exceeds that of the neutral gas and the UC\hii regions are still expanding. 

  \item The percentage of the dust cores that are associated with \hii regions are 33\%, 73\%, 4\%, and 1\% for Sgr\,B2(N), Sgr\,B2(M), Sgr\,B2(S), and Sgr\,B2(DS), respectively.
  Among all the dust cores in Sgr\,B2(DS), two-thirds are associated with outflows that are traced by SiO(5--4).
  The dust cores in both of Sgr\,B2(M) and Sgr\,B2(N) are associated with two 6.7\,GHz \ce{CH3OH} masers, while the dust cores in Sgr\,B2(DS) have no associated maser.
  Based on these findings, we suggest that the dust cores in Sgr\,B2(M) are more evolved than those in Sgr\,B2(N).
  The dust cores in Sgr\,B2(DS) are younger than those in Sgr\,B2(M) or Sgr\,B2(N).
\end{itemize}

\meng{In this work, we calculated the physical properties of the UC\hii regions and their surrounding neutral gas in Sgr\,B2. We found that the pressure of the UC\hii regions and the dense gas surrounding them are comparable. We also characterized the evolutionary stages of these UC\hii regions and obtained their minimum expansion timescales. 
}

\begin{acknowledgements}
  FM, ASM, PS, ASchw research is carried out within the Collaborative Research Centre 956, sub-projects A6 and C3, funded by the Deutsche Forschungsgemeinschaft (DFG) - project ID 184018867. AG acknowledges support from the NSF via AST 2008101 and CAREER 2142300. This paper makes use of the following ALMA data: 2013.1.00269.S and 2017.1.00114.S. ALMA is a partnership of ESO (representing its member states), NSF (USA) and NINS (Japan), together with NRC (Canada), MOST and ASIAA (Taiwan), and KASI (Republic of Korea), in cooperation with the Republic of Chile. The Joint ALMA Observatory is operated by ESO, AUI/NRAO and NAOJ. This research made use of Astropy,\footnote{http://www.astropy.org} a community-developed core Python package for Astronomy \citep{Astropy-Collaboration:2013aa,Astropy-Collaboration:2018aa}.
\end{acknowledgements}

\bibliographystyle{aa} 
\bibliography{ref}

\begin{appendix}

\section{RMS maps} 
\label{sec:rms_maps}

\begin{figure}[ht]
    \begin{center}
      \includegraphics[width=0.32\textwidth]{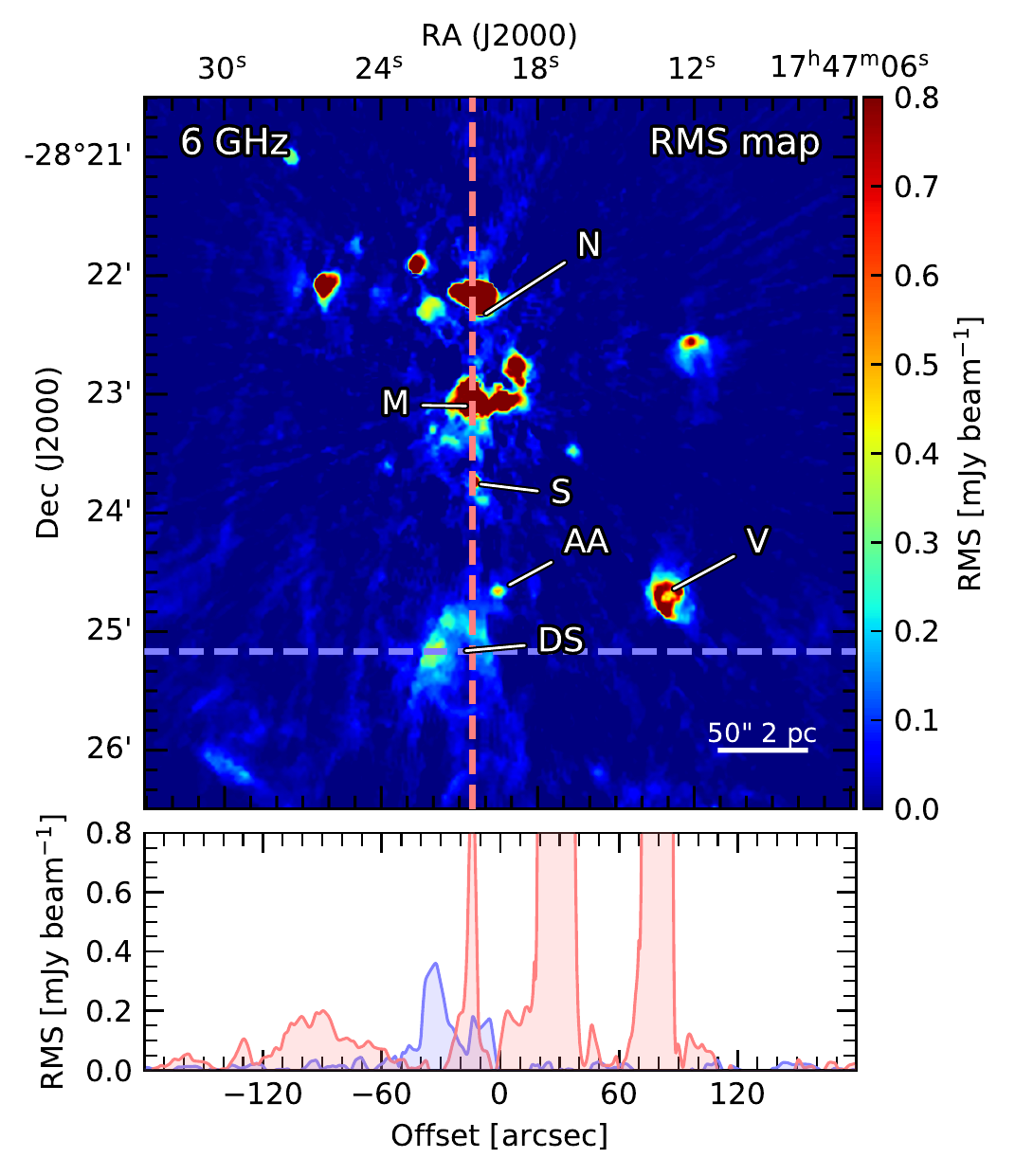}
      \includegraphics[width=0.32\textwidth]{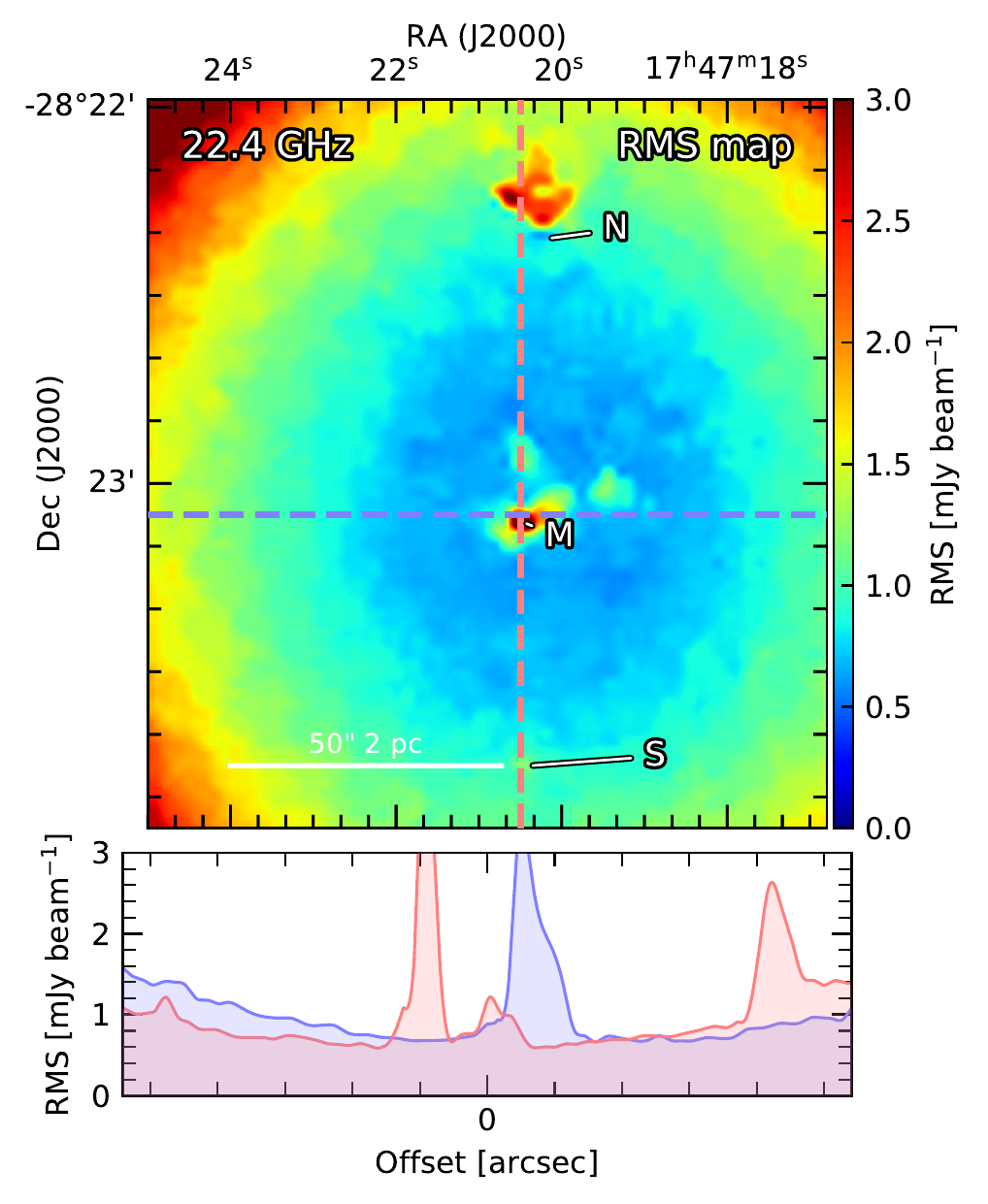}
      \includegraphics[width=0.32\textwidth]{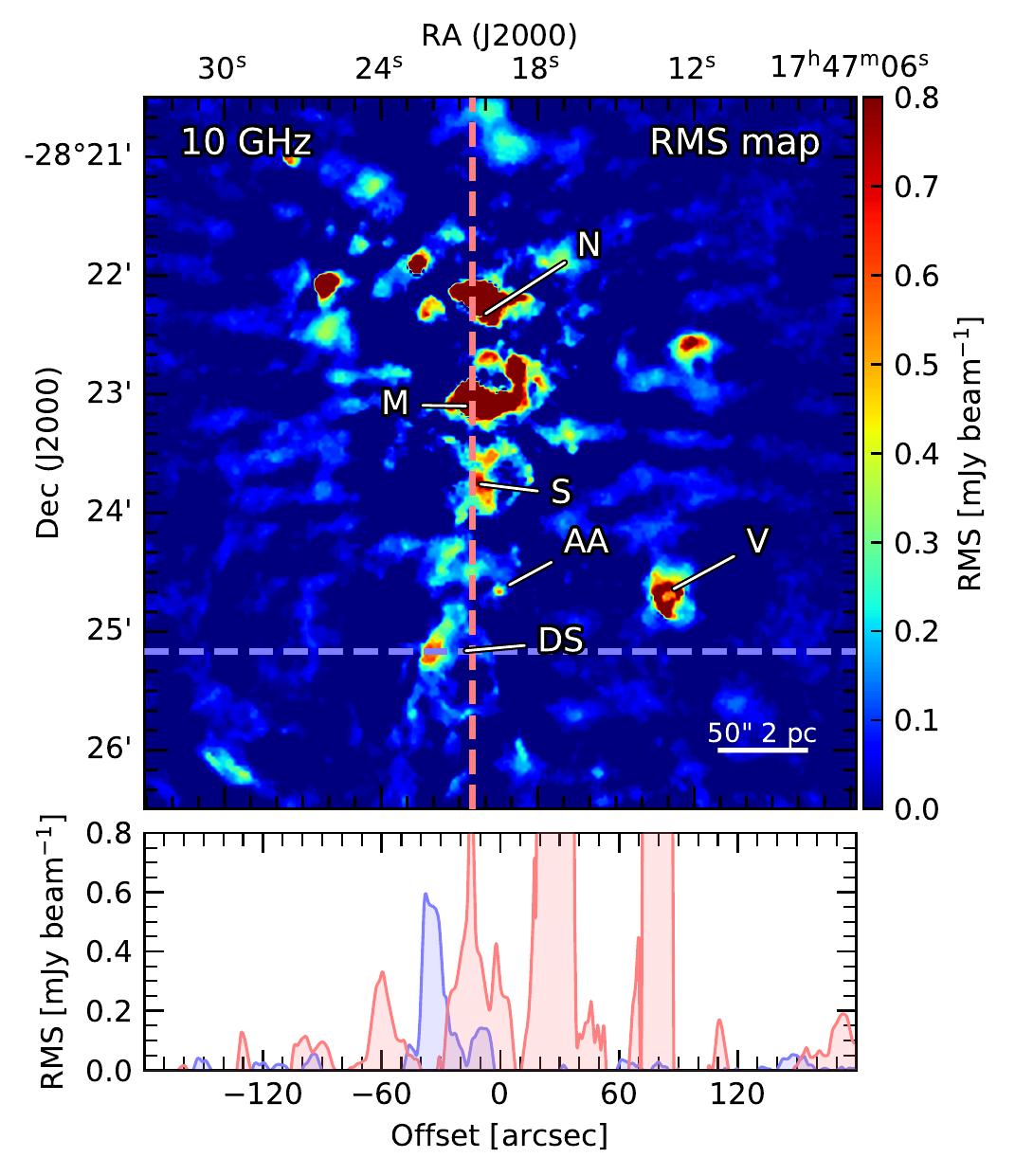}
      \caption[RMS maps]{RMS maps of 6\,GHz, 22.4\,GHz, and 96\,GHz.\LEt{ from left to right?   I read 10 GHz in the right image. ("respectively" not appropriate here) } The resolutions are given in Sect.\,\ref{sec:observations_and_data_reduction}. Each image has two profile cuts, in blue and red. To better show the features and maxima of the rms of each image, the offset is different for each of the three images. The offset of each profile increases from left to right and from bottom to top. The profile is below each image.
       }
      \label{f:rmsmaps}
    \end{center}
  \end{figure}

  The rms maps were generated using SExtractor \citep{Bertin:1996aa}.
  Although the core identification was performed by eye, SExtractor produces rms maps as side-products of automated core identification.
  The rms maps are shown in Fig.\,\ref{f:rmsmaps}.

\section{Tables} 

  \begin{table*}
    \begin{small}
    \caption{Observed parameters}
    \label{t:coreparams}
\centering{
    \begin{tabular}{lrrrrrrrrr}
    \toprule
    \hline \noalign{\smallskip}
    \# &
    RA & 
    DEC &
    $r_{\rm obs6}$ &
    $S_{6}$ &
    $r_{\rm obs22}$ &
    $S_{22}$ &
    $r_{\rm obs96}$ &
    $S_{96}$ &
    $\alpha_{6-22}$\\
    &
    17:47:- - & 
    $-$28:- - &
    arcsec&
    mJy &
    arcsec&
    mJy &
    arcsec&
    mJy  &
    \\
    \hline \noalign{\smallskip}
1$^{* }$&23.355&25:33.95&0.70&$2.4 \pm 0.1$&...&...&0.35&$0.3 \pm 0.2$                       & ...                     \\ 
2$^{a }$&19.485&24:39.75&1.45&$64.9 \pm 0.3$&...&...&1.20&$60.9 \pm 0.2$                     & ...                       \\ 
3$^{* }$&20.235&23:44.95&0.60&$7.0 \pm 0.1$&0.60&$14.7 \pm 0.7$&0.60&$7.9 \pm 0.2$           & $0.56  \pm 0.26$                            \\ 
4$^{a }$&20.428&23:44.25&2.10&$363.1 \pm 0.8$&1.15&$820.5 \pm 1.6$&1.45&$1107.6 \pm 0.5$     & $0.62  \pm 0.27$                                  \\ 
5$^{a }$&20.042&23:18.00&1.65&$123.8 \pm 0.2$&1.10&$160.4 \pm 0.9$&1.05&$135.5 \pm 0.2$      & $0.20  \pm 0.09$                                 \\ 
6$^{a }$&20.049&23:12.60&1.20&$68.1 \pm 0.2$&0.65&$126.1 \pm 0.7$&0.80&$124.8 \pm 0.2$       & $0.47  \pm 0.21$                                \\ 
7$^{b }$&19.765&23:09.90&0.75&$15.6 \pm 0.1$&0.50&$22.8 \pm 0.7$&0.40&$18.3 \pm 0.2$         & $0.29  \pm 0.13$                              \\ 
8$^{* }$&20.015&23:08.85&0.55&$28.3 \pm 0.3$&0.55&$64.2 \pm 0.7$&0.55&$52.5 \pm 0.6$         & $0.62  \pm 0.27$                              \\ 
9$^{a }$&20.106&23:08.45&0.90&$96.9 \pm 0.5$&0.70&$274.6 \pm 0.7$&0.75&$266.3 \pm 1.1$       & $0.79  \pm 0.35$                                \\ 
10$^{b }$&20.315&23:08.00&0.40&$9.2 \pm 0.6$&0.30&$25.5 \pm 0.7$&0.45&$37.3 \pm 0.9$         & $0.77  \pm 0.35$                              \\ 
11$^{b }$&19.799&23:06.70&0.80&$31.3 \pm 0.3$&0.60&$63.3 \pm 0.7$&0.55&$39.0 \pm 0.3$        & $0.53  \pm 0.24$                               \\ 
12$^{b }$&20.148&23:06.60&0.35&$7.0 \pm 0.4$&0.30&$14.1 \pm 0.7$&0.35&$44.0 \pm 1.1$         & $0.53  \pm 0.25$                              \\ 
13$^{b }$&20.193&23:06.55&0.30&$4.2 \pm 0.4$&0.30&$20.3 \pm 0.7$&0.40&$132.9 \pm 1.6$        & $1.20  \pm 0.54$                               \\ 
14$^{b }$&20.246&23:06.10&0.65&$30.9 \pm 1.0$&0.45&$28.4 \pm 1.1$&0.45&$46.4 \pm 3.0$        & $-0.06 \pm 0.07$                               \\ 
15$^{b }$&20.133&23:06.05&0.35&$6.9 \pm 0.4$&0.25&$12.7 \pm 0.7$&0.30&$59.0 \pm 1.0$         & $0.46  \pm 0.23$                              \\ 
16$^{b }$&20.174&23:05.80&0.40&$11.8 \pm 0.6$&0.30&$45.9 \pm 0.8$&0.50&$267.1 \pm 1.9$       & $1.03  \pm 0.46$                                \\ 
17$^{b }$&19.693&23:05.65&0.65&$15.9 \pm 0.3$&0.40&$20.2 \pm 0.7$&0.40&$13.6 \pm 0.2$        & $0.18  \pm 0.10$                               \\ 
18$^{b }$&20.072&23:05.10&0.45&$12.7 \pm 0.4$&0.30&$22.0 \pm 0.7$&0.40&$45.5 \pm 1.5$        & $0.42  \pm 0.19$                               \\ 
19$^{ab}$&20.159&23:04.70&0.65&$114.6 \pm 0.8$&0.60&$932.9 \pm 1.3$&0.70&$3194.7 \pm 2.8$    & $1.59  \pm 0.70$                                   \\ 
20$^{ab}$&20.303&23:04.40&0.55&$27.5 \pm 1.0$&...&...&0.55&$52.3 \pm 2.7$                    & ...                     \\ 
21$^{ab}$&20.231&23:04.20&0.55&$58.4 \pm 0.8$&0.40&$245.2 \pm 0.9$&0.40&$528.6 \pm 1.7$      & $1.09  \pm 0.48$                                 \\ 
22$^{a }$&20.121&23:03.90&0.35&$41.5 \pm 0.3$&0.45&$487.2 \pm 0.9$&0.50&$1353.7 \pm 2.0$     & $1.87  \pm 0.82$                                  \\ 
23$^{a }$&20.174&23:03.50&0.45&$47.1 \pm 0.4$&0.40&$225.3 \pm 0.8$&0.40&$517.8 \pm 1.5$      & $1.19  \pm 0.52$                                 \\ 
24$^{ab}$&20.106&23:03.40&0.30&$15.5 \pm 0.3$&...&...&0.40&$450.9 \pm 1.6$                   & ...                       \\ 
25$^{ab}$&20.246&23:03.25&0.25&$4.3 \pm 0.3$&0.25&$12.9 \pm 0.7$&0.30&$80.9 \pm 1.1$         & $0.83  \pm 0.38$                              \\ 
26$^{b }$&20.106&23:02.90&0.30&$8.1 \pm 0.2$&0.30&$10.3 \pm 0.7$&0.35&$47.2 \pm 1.2$         & $0.18  \pm 0.12$                              \\ 
27$^{ab}$&20.277&23:02.85&0.45&$24.0 \pm 0.4$&0.45&$198.2 \pm 0.8$&0.50&$387.9 \pm 1.5$      & $1.60  \pm 0.70$                                 \\ 
28$^{ab}$&19.902&23:02.80&0.85&$100.7 \pm 0.3$&0.60&$323.9 \pm 0.8$&0.85&$360.2 \pm 0.6$     & $0.89  \pm 0.39$                                  \\ 
29$^{b }$&19.992&23:02.65&0.45&$13.1 \pm 0.2$&0.40&$17.0 \pm 0.7$&0.40&$19.8 \pm 0.7$        & $0.20  \pm 0.10$                               \\ 
30$^{b }$&20.269&23:02.05&0.45&$13.0 \pm 0.3$&0.30&$16.6 \pm 0.7$&0.45&$44.4 \pm 1.0$        & $0.19  \pm 0.10$                               \\ 
31$^{ab}$&20.125&23:02.00&0.95&$56.3 \pm 0.6$&0.55&$62.1 \pm 0.9$&0.55&$73.3 \pm 1.3$        & $0.07  \pm 0.04$                               \\ 
32$^{b }$&19.898&23:01.90&0.35&$11.3 \pm 0.1$&0.30&$18.5 \pm 0.7$&0.45&$40.5 \pm 0.2$        & $0.37  \pm 0.17$                               \\ 
33$^{b }$&19.860&23:01.35&0.45&$10.4 \pm 0.1$&0.30&$19.8 \pm 0.7$&0.45&$24.1 \pm 0.2$        & $0.49  \pm 0.22$                               \\ 
34$^{b }$&20.125&22:60.00&0.70&$22.7 \pm 0.3$&...&...&0.75&$41.6 \pm 0.5$                    &  ...                      \\ 
35$^{* }$&20.087&22:57.05&0.40&$1.6 \pm 0.1$&...&...&0.50&$5.3 \pm 0.2$                      &  ...                     \\ 
36$^{a }$&19.598&22:56.25&0.75&$30.9 \pm 0.2$&0.70&$38.7 \pm 0.7$&0.55&$26.7 \pm 0.2$        & $0.17  \pm 0.08$                               \\ 
37$^{a }$&20.186&22:56.25&0.40&$3.1 \pm 0.1$&...&...&0.40&$3.9 \pm 0.2$                      & ...                       \\ 
38$^{a }$&19.515&22:56.15&0.60&$30.3 \pm 0.2$&0.40&$44.2 \pm 0.7$&0.55&$49.6 \pm 0.2$        & $0.29  \pm 0.13$                               \\ 
39$^{* }$&19.939&22:55.50&0.50&$1.4 \pm 0.1$&...&...&0.50&$3.1 \pm 0.2$                      & ...                         \\ 
40$^{a }$&19.526&22:55.30&0.65&$36.6 \pm 0.1$&0.40&$43.0 \pm 0.7$&0.55&$45.6 \pm 0.2$        & $0.12  \pm 0.06$                               \\ 
41$^{a }$&18.629&22:54.30&1.25&$73.9 \pm 0.5$&...&...&1.15&$84.5 \pm 0.3$                    & ...                          \\ 
42$^{* }$&20.000&22:47.45&0.70&$5.4 \pm 0.1$&...&...&0.70&$6.0 \pm 0.2$                      & ...                        \\ 
43$^{* }$&19.886&22:47.40&0.60&$3.1 \pm 0.1$&...&...&0.60&$5.4 \pm 0.2$                      & ...                        \\ 
44$^{b }$&20.030&22:41.15&0.50&$5.0 \pm 0.1$&0.35&$32.2 \pm 0.7$&0.55&$79.0 \pm 0.2$         & $1.41  \pm 0.62$                              \\ 
45$^{b }$&19.489&22:26.40&0.65&$6.1 \pm 0.1$&...&...&0.80&$11.7 \pm 0.2$                     & ...                          \\ 
46$^{a }$&19.803&22:20.70&1.10&$79.1 \pm 0.1$&0.65&$227.5 \pm 0.8$&0.50&$141.0 \pm 0.4$      & $0.80  \pm 0.35$                                 \\ 
47$^{* }$&19.417&22:19.75&0.60&$3.0 \pm 0.1$&...&...&0.60&$7.3 \pm 0.2$                      & ...                       \\ 
48$^{a }$&19.871&22:18.30&0.45&$7.8 \pm 0.1$&0.30&$43.5 \pm 0.7$&0.95&$1401.0 \pm 5.4$       & $1.30  \pm 0.57$                                \\ 
49$^{a }$&19.898&22:17.00&0.65&$28.2 \pm 0.4$&0.45&$184.2 \pm 0.7$&0.50&$290.9 \pm 3.1$      & $1.42  \pm 0.63$                                 \\ 
50$^{b }$&18.102&22:06.85&0.50&$6.4 \pm 0.1$&0.25&$8.9 \pm 0.7$&0.50&$9.7 \pm 0.2$           & $0.25  \pm 0.15$                            \\ 
51$^{a }$&19.996&22:04.65&1.20&$162.6 \pm 0.9$&0.80&$411.3 \pm 1.5$&1.00&$271.4 \pm 0.6$     & $0.70  \pm 0.31$                                  \\ 
52$^{a }$&17.337&22:03.60&0.75&$16.5 \pm 0.1$&0.45&$29.8 \pm 1.1$&0.65&$19.8 \pm 0.2$        & $0.45  \pm 0.20$                               \\ 
53$^{b }$&23.049&21:55.55&0.85&$23.6 \pm 0.1$&...&...&0.90&$30.3 \pm 0.2$                    & ...                       \\ 
54$^{a }$&24.830&21:44.34&1.95&$20.6 \pm 0.1$&...&...&0.45&$4.1 \pm 0.2$                     & ...                      \\

    \bottomrule      
      \end{tabular}
    {\\
    ${a}$: Associated with the \hii regions identified by \cite{Benson:1984wr}.\\
    ${b}$: Associated with the \hii regions identified by \cite{Gaume:1995aa}.\\
    ${*}$: Newly identified in this study.
    }

 }
\end{small}
    \end{table*}

 \begin{table*}
    \begin{small}
    \caption{Derived parameters}
    \label{t:coreparams_derived}
    \centering{
    \begin{tabular}{rrrrrrrr}
    \toprule
    \hline \noalign{\smallskip}
    \# &
    $r_{\rm calc}$ & 
    ${\rm EM}_{\rm calc}$ &
    $\log_{10}(\dot{N}_{\rm Ly}/{\rm s^{-1}})$ &
    $n_{\rm e}$ &
    $n_{\rm H_2}$ &
    $r_i$ &
    $t$\\
    &
    $\times 10^{-3}$\,pc & 
    $\times 10^7\,{\rm pc\,cm^{-6}}$ &
    &
    $\times 10^4\,{\rm cm^{-3}}$&
    $\times 10^6\,{\rm cm^{-3}}$&
    $\times 10^{-3}$\,pc &
    $\times 10^4\,{\rm yrs}$\\

    \hline \noalign{\smallskip}
    1&$^*$26.73&$^*$0.24&$^*$46.17&$^*$0.94&...&$^{\dagger}$$^*$0.106&$^{\dagger}$$^*$9.3\\ 
2&$^*$57.40&$^*$1.49&$^*$47.64&$^*$1.61&0.22&$^*$0.324&$^*$15.2\\ 
3&6.51&30.34&47.05&21.59&...&$^{\dagger}$0.207&$^{\dagger}$0.5\\ 
4&46.19&33.91&48.80&8.57&4.69&0.794&5.3\\ 
5&33.59&11.90&48.07&5.95&...&$^{\dagger}$0.453&$^{\dagger}$4.6\\ 
6&20.84&25.06&47.98&10.97&0.77&0.422&2.1\\ 
7&10.95&16.11&47.23&12.13&...&$^{\dagger}$0.237&$^{\dagger}$1.1\\ 
8&12.90&33.98&47.70&16.23&...&$^{\dagger}$0.339&$^{\dagger}$1.1\\ 
9&23.30&45.84&48.34&14.03&0.33&0.556&2.1\\ 
10&7.21&44.38&47.30&24.82&5.94&0.251&0.5\\ 
11&13.85&28.75&47.68&14.41&...&$^{\dagger}$0.337&$^{\dagger}$1.2\\ 
12&6.54&28.68&47.03&20.95&30.27&0.097&0.8\\ 
13&4.79&88.33&47.25&42.92&72.90&0.064&0.7\\ 
14&40.83&1.39&47.31&1.85&9.97&0.253&10.1\\ 
15&6.66&24.78&46.98&19.29&73.46&0.052&1.4\\ 
16&8.02&68.02&47.58&29.12&73.96&0.082&1.4\\ 
17&12.22&11.28&47.17&9.61&...&$^{\dagger}$0.227&$^{\dagger}$1.3\\ 
18&9.17&22.45&47.22&15.65&16.66&0.168&1.0\\ 
19&24.94&183.33&48.99&27.11&244.93&0.109&8.0\\ 
20&$^*$20.36&$^*$5.88&$^*$47.33&$^*$5.37&6.66&$^*$0.257&$^*$3.0\\ 
21&17.85&74.51&48.32&20.43&182.20&0.079&5.7\\ 
22&15.02&384.53&48.86&50.60&157.61&0.132&2.9\\ 
23&16.00&87.68&48.29&23.41&184.11&0.077&4.8\\ 
24&$^*$8.63&$^*$402.88&$^*$48.39&$^*$68.31&95.39&$^*$0.129&$^*$1.1\\ 
25&4.90&48.99&47.01&31.62&106.05&0.042&1.0\\ 
26&8.70&11.39&46.88&11.44&37.11&0.076&1.7\\ 
27&11.41&188.10&48.32&40.61&44.63&0.202&1.3\\ 
28&23.58&53.78&48.42&15.10&3.03&0.591&2.0\\ 
29&10.91&11.93&47.10&10.45&3.11&0.214&1.1\\ 
30&11.05&11.35&47.09&10.14&13.64&0.173&1.4\\ 
31&27.61&6.74&47.66&4.94&4.72&0.329&4.2\\ 
32&8.87&20.04&47.14&15.03&10.94&0.209&0.8\\ 
33&8.06&26.49&47.18&18.13&2.72&0.228&0.6\\ 
34&$^*$26.73&$^*$2.49&$^*$47.20&$^*$3.05&2.25&$^*$0.231&$^*$5.1\\ 
35&$^*$13.66&$^*$0.64&$^*$46.02&$^*$2.16&1.35&$^*$0.094&$^*$3.1\\ 
36&17.30&10.80&47.45&7.90&...&$^{\dagger}$0.282&$^{\dagger}$2.1\\ 
37&$^*$13.66&$^*$1.25&$^*$46.31&$^*$3.02&0.92&$^*$0.117&$^*$2.6\\ 
38&15.30&15.95&47.52&10.21&2.51&0.296&1.6\\ 
39&$^*$18.18&$^*$0.31&$^*$45.95&$^*$1.30&0.67&$^*$0.089&$^*$5.4\\ 
40&20.22&8.74&47.50&6.58&1.93&0.292&2.7\\ 
41&$^*$49.30&$^*$2.37&$^*$47.71&$^*$2.19&0.64&$^*$0.342&$^*$11.2\\ 
42&$^*$26.73&$^*$0.56&$^*$46.54&$^*$1.44&0.21&$^*$0.140&$^*$7.5\\ 
43&$^*$22.51&$^*$0.44&$^*$46.30&$^*$1.41&0.58&$^*$0.116&$^*$6.4\\ 
44&18.18&1.12&46.51&2.48&18.89&0.090&5.3\\ 
45&$^*$24.63&$^*$0.74&$^*$46.60&$^*$1.73&0.56&$^*$0.146&$^*$6.3\\ 
46&21.03&46.73&48.26&14.91&...&$^{\dagger}$0.522&$^{\dagger}$1.8\\ 
47&$^*$22.51&$^*$0.43&$^*$46.28&$^*$1.38&0.97&$^*$0.115&$^*$6.4\\ 
48&6.51&106.80&47.59&40.51&66.06&0.089&0.9\\ 
49&12.39&131.78&48.24&32.62&26.56&0.269&1.2\\ 
50&7.28&14.09&46.82&13.92&0.61&0.173&0.7\\ 
51&30.47&39.57&48.51&11.39&...&$^{\dagger}$0.633&$^{\dagger}$3.0\\ 
52&10.35&23.94&47.35&15.21&...&$^{\dagger}$0.261&$^{\dagger}$0.9\\ 
53&$^*$32.96&$^*$1.65&$^*$47.20&$^*$2.23&0.65&$^*$0.232&$^*$7.4\\ 
54&$^*$77.55&$^*$0.25&$^*$47.12&$^*$0.56&...&$^{\dagger}$$^*$0.218&$^{\dagger}$$^*$34.7\\

    \bottomrule
              
      \end{tabular}
      }
      \tablefoot{$*$: Sources that are without detection in 22.4\,GHz. Parameters are derived from 6\,GHz flux and the deconvolved core size. $\dagger$: No detection in 96\,GHz. Parameters are derived assuming $n_{\rm H_2} = 2\times10^{7}\,{\rm cm^{-3}}$
}
\end{small}
    \end{table*}

 \begin{table*}
    \begin{small}
    \caption{Which cores in \citet{Ginsburg:2018wo} are associates with outflows.}
    \label{t:g18_outflow}
    \centering{
    \begin{tabular}{rl|rl|rl|rl|rl|rl}
    \toprule
    \hline \noalign{\smallskip}
    ID & Outflow &
    ID & Outflow &
    ID & Outflow &
    ID & Outflow &
    ID & Outflow &
    ID & Outflow \\
\hline \noalign{\smallskip}
 1&   Y &  47 &  -- &  97  & -- &  143 &  N &  202 & -- &           258 &  Y \\
 2&   Y &  48 &  -- &  98  & -- &  144 &  N &  203 & -- &           259 &  Y \\
 3&   Y &  49 &   Y &  99  & -- &  145 & -- &  204 & -- &           260 &  Y \\
 4&   Y &  50 &   Y &  100 & -- &  146 & -- &  205 & -- &           261 &  N \\
 5&   Y &  51 &   Y &  102 & -- &  147 &  N &  206 & -- &           262 &  N \\
 6&   N &  52 &  -- &  103 & -- &  148 & -- &  207 &  Y &           263 &  N \\
 7&   Y &  53 &  -- &  104 & -- &  149 & -- &  208 &  N &           266 &  N \\
 8&   Y &  54 &   Y &  105 & -- &  150 & -- &  209 &  N &           267 &  N \\
 9&   N &  55 &  -- &  106 & -- &  153 & -- &  210 &  Y &           268 &  N \\
10&   N &  56 &  -- &  107 & -- &  154 & -- &  211 &  Y &           269 &  N \\
11&   N &  57 &  -- &  108 & -- &  155 & -- &  212 &  Y &           270 &  N \\
12&   Y &  58 &  -- &  109 & -- &  156 & -- &  213 & -- &           271 &  N \\
13&  -- &  59 &  -- &  110 & -- &  157 & -- &  214 & -- &        101\_X & -- \\
14&   N &  60 &  -- &  111 & -- &  158 &  Y &  215 & -- &        175\_G & -- \\
15&  -- &  61 &  -- &  112 & -- &  159 &  Y &  216 & -- &        177\_B & -- \\
16&   Y &  62 &  -- &  113 & -- &  160 &  Y &  217 & -- &        180\_E & -- \\
17&   Y &  63 &  -- &  114 & -- &  161 &  Y &  218 & -- &        181\_D & -- \\
18&   Y &  64 &  -- &  115 & -- &  162 &  Y &  219 & -- &        182\_Y & -- \\
19&   Y &  65 &  -- &  116 &  N &  163 &  Y &  220 & -- &        244\_C & -- \\
20&   Y &  66 &  -- &  117 &  N &  164 & -- &  221 & -- &        265\_H &  N \\
21&   N &  67 &  -- &  118 &  N &  165 & -- &  222 & -- &       171\_K3 & -- \\
22&   Y &  68 &  -- &  119 & -- &  166 & -- &  223 & -- &       172\_K2 & -- \\
23&   Y &  69 &  -- &  120 &  Y &  167 & -- &  224 & -- &       173\_K1 & -- \\
24&   Y &  70 &   N &  121 &  N &  168 & -- &  226 & -- &       174\_f3 & -- \\
25&   Y &  71 &   N &  122 &  N &  169 & -- &  227 & -- &       176\_f1 & -- \\
26&  -- &  72 &   N &  123 &  N &  170 & -- &  228 & -- &       234\_f4 & -- \\
27&  -- &  73 &   N &  124 &  N &  183 &  N &  229 & -- &       235\_f2 & -- \\
28&  -- &  74 &   N &  125 &  N &  184 &  N &  230 & -- &       245\_A2 & -- \\
29&  -- &  75 &   N &  126 &  N &  185 &  N &  231 & -- &       264\_k4 & -- \\
30&  -- &  76 &   N &  127 &  N &  186 &  N &  232 & -- &     86\_B9.96 & -- \\
31&  -- &  77 &   N &  128 &  N &  187 &  N &  238 & -- &     87\_B9.99 & -- \\
32&  -- &  78 &   N &  129 &  N &  188 &  N &  239 & -- &     90\_B9.89 & -- \\
33&  -- &  79 &   N &  130 & -- &  189 &  N &  243 & -- &    92\_I10.52 & -- \\
34&   Y &  80 &   N &  131 & -- &  190 &  N &  246 & -- &    96\_Z10.24 & -- \\
35&   Y &  81 &   N &  132 & -- &  191 &  N &  247 & -- &   151\_B10.06 & -- \\
36&   Y &  82 &   N &  133 & -- &  192 &  N &  248 &  N &   152\_f10.32 & -- \\
37&   N &  83 &   N &  134 & -- &  193 &  N &  249 &  N &   178\_f10.37 & -- \\
38&   Y &  84 &  -- &  135 & -- &  194 &  N &  250 & -- &   179\_f10.38 & -- \\
39&  -- &  85 &  -- &  136 & -- &  195 &  Y &  251 & -- &   237\_G10.44 & -- \\
40&  -- &  88 &  -- &  137 & -- &  196 &  Y &  252 & -- &   241\_f10.30 & -- \\
41&  -- &  89 &  -- &  138 & -- &  197 &  Y &  253 & -- &  225\_f10.33b & -- \\
42&  -- &  91 &  -- &  139 & -- &  198 &  Y &  254 &  N &  233\_f10.27b & -- \\
43&  -- &  93 &  -- &  140 &  Y &  199 & -- &  255 &  N &  236\_f10.303 & -- \\
44&  -- &  94 &  -- &  141 &  Y &  200 & -- &  256 &  Y &  240\_f10.44b & -- \\
45&   Y &  95 &  -- &  142 & -- &  201 & -- &  257 &  Y &  242\_f10.318 & -- \\
46&   Y &     &     &      &    &      &    &      &    &               &    \\
    \bottomrule
              
      \end{tabular}
}
      \tablefoot{The IDs of the dust cores are as in Table~{3} in \citet{Ginsburg:2018wo}. Cores that are not covered by the SiO image (Fig.~\ref{f:small.sio_mom1}) are marked   ``--''. Cores that are covered by the SiO image and associated or not associated with outflows are flagged  Y or N, respectively. }

\end{small}
   
    \end{table*}

\section{SiO Images}

          \begin{sidewaysfigure*}
    \begin{center}
    \includegraphics[width=0.75\textwidth]{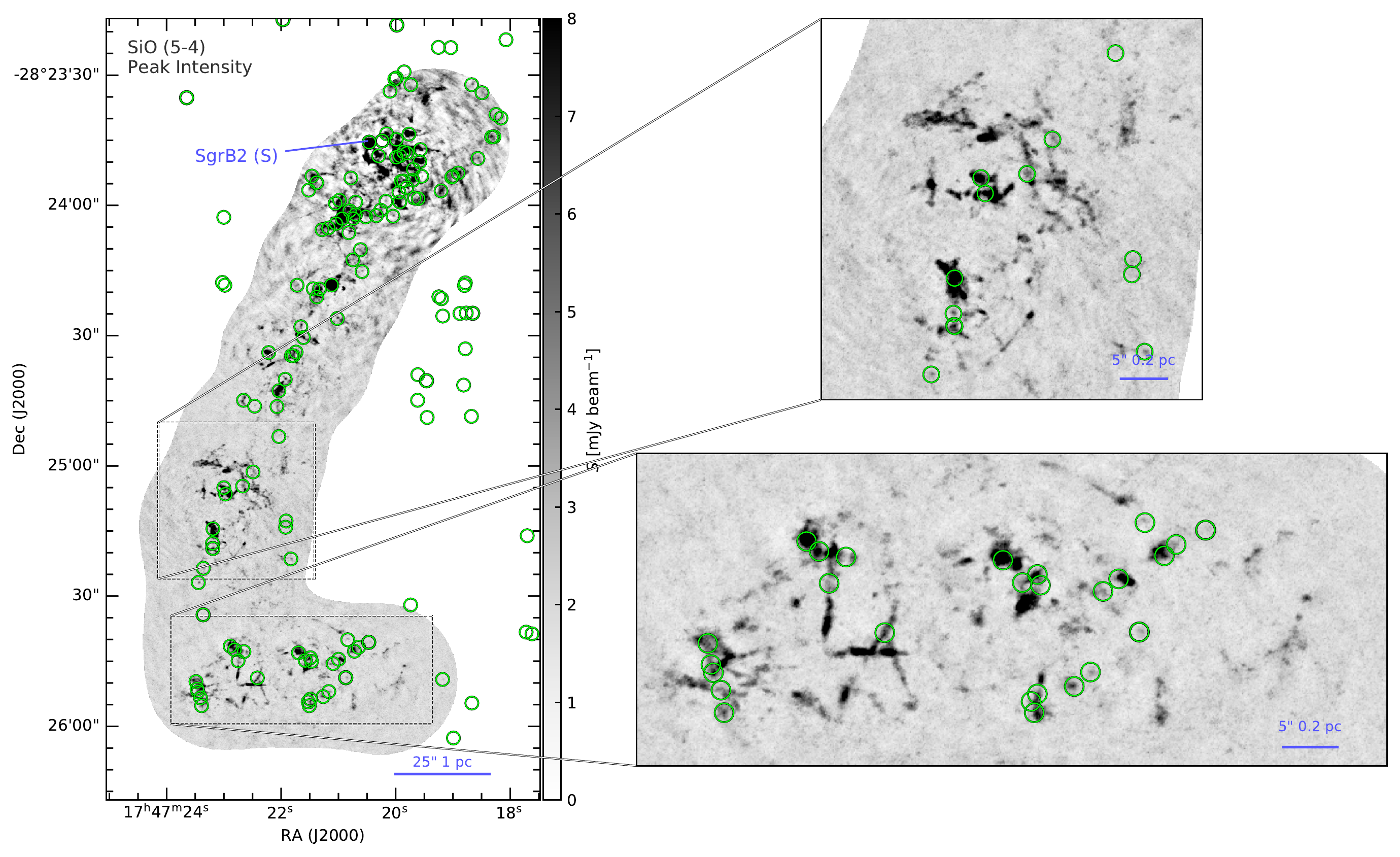}        
    \caption[Peak intensity of SiO (5-4)]{Peak intensity map of SiO (5-4) line. The zoomed-in plots are the regions in Sgr\,B2(DS)  with abundant outflows. The resolution of the image is 0.35\arcsec $\times$ 0.24\arcsec, with P.A. of $-80^{\circ}$. The dust cores identified by \citet{Ginsburg:2018wo} are shown  as green circles, whose size is not scaled to the size of the cores.}
    \label{f:small.sio_peak}
    \end{center}
\end{sidewaysfigure*}

  \begin{sidewaysfigure*}
    \begin{center}
    \includegraphics[width=0.75\textwidth]{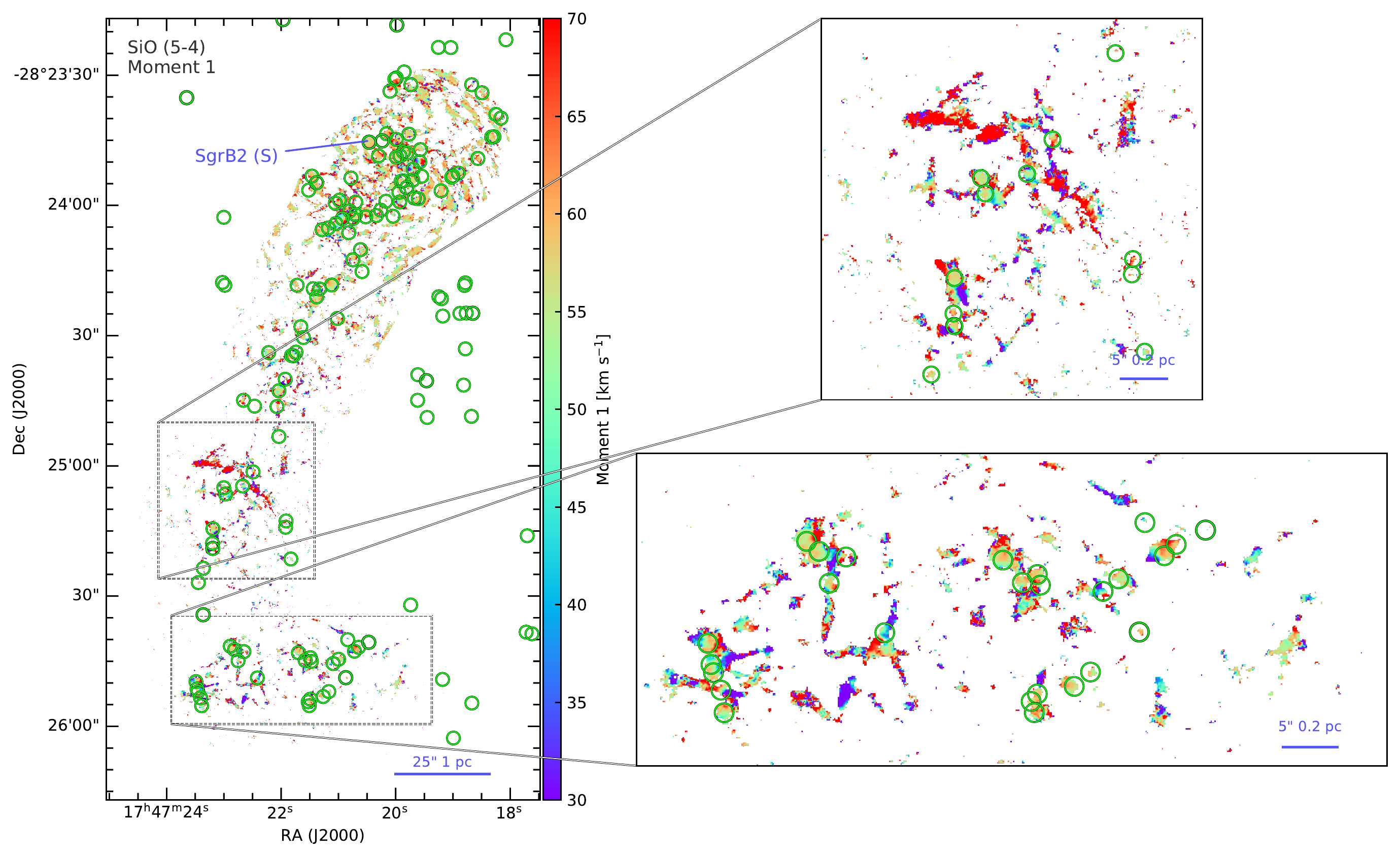}        
    \caption[Moment 1 of SiO (5-4)]{Moment 1 map of SiO (5-4) line. The zoomed-in plots are the regions in Sgr\,B2(DS)   with abundant outflows. The resolution of the image is 0.35\arcsec $\times$ 0.24\arcsec, with P.A. of $-80^{\circ}$. The dust cores identified by \citet{Ginsburg:2018wo} are shown as green circles, whose size is not scaled  to the size of the cores.}
    \label{f:small.sio_mom1}
    \end{center}
\end{sidewaysfigure*}

\end{appendix}

\end{document}